\documentclass[number,sort&compress,twocolumn,5p]{elsarticle}
\usepackage[binary-units=true,separate-uncertainty=true]{siunitx}
\usepackage{lineno,hyperref}
\usepackage{tikz}
\usepackage{amssymb,amsmath,mathtools}
\usepackage{caption}
\usepackage{xspace}
\usepackage{booktabs}
\usepackage{siunitx}
\usepackage{textcomp}
\usepackage[frozencache]{minted}
\usepackage[percent]{overpic}
\usepackage[list=true,listformat=simple,margin=10pt]{subcaption}
\usepackage{hepnames}

\journal{Nucl. Instr. Meth. A}

\newcommand{\apsq}{\texorpdfstring{\ensuremath{\mathrm{Allpix}^2}}{Allpix\textasciicircum 2}\xspace}

\newcommand{\SIERRS}[4]{\ensuremath{\num{#1}\,^{#2}_{#3}\,\text{(syst)}\,\si{#4}}}
\newcommand{\SIERRA}[5]{\ensuremath{\num{#1}\pm\num{#2}\,\text{(stat)}\,^{#3}_{#4}\,\text{(syst)}\,\si{#5}}}

\begin{document}
\begin{frontmatter}
  \title{Combining TCAD and Monte Carlo Methods to Simulate CMOS Pixel Sensors with a Small Collection Electrode using the Allpix$^2$ Framework}

\author[cern]{D.~Dannheim}
\author[cern]{K.~Dort\fnref{ugiessen}}
\author[cern]{D.~Hynds\fnref{nikhef}}
\author[cern]{M.~Munker}
\author[cern]{A.~N\"urnberg\fnref{kit}}
\author[cern]{W.~Snoeys}
\author[cern]{S.~Spannagel\corref{corr}}
\ead{simon.spannagel@cern.ch}

\address[cern]{CERN, Geneva, Switzerland}

\cortext[corr]{Corresponding author}

\fntext[ugiessen]{Also at University of Giessen, Germany}
\fntext[nikhef]{Now at Nikhef, Amsterdam, Netherlands}
\fntext[kit]{Now at KIT, Karlsruhe, Germany}

\begin{abstract}
Combining electrostatic field simulations with Monte Carlo methods enables realistic modeling of the detector response for novel monolithic silicon detectors with strongly non-linear electric fields.
Both the precise field description and the inclusion of Landau fluctuations and production of secondary particles in the sensor are crucial ingredients for the understanding and reproduction of detector characteristics.

In this paper, a CMOS pixel sensor with small collection electrode design, implemented in a high-resistivity epitaxial layer, is simulated by integrating a detailed electric field model from finite element TCAD into a Monte Carlo based simulation with the \apsq framework.
The simulation results are compared to data recorded in test-beam measurements and very good agreement is found for various quantities such as cluster size, spatial resolution and efficiency.
Furthermore, the observables are studied as a function of the intra-pixel incidence position to enable a detailed comparison with the detector behavior observed in data.

The validation of such simulations is fundamental for modeling the detector response and for predicting the performance of future prototype designs.
Moreover, visualization plots extracted from the charge carrier drift model of the framework can aid in understanding the charge propagation behavior in different regions of the sensor.

\end{abstract}


\end{frontmatter}

\tableofcontents

\section{Introduction}
\label{sec:introduction}
Integrated monolithic CMOS technologies with small collection electrodes~\cite{Snoeys} are emerging technologies enabling advances in the design of next-generation high-performance silicon vertex and tracking detectors for high-energy physics.  These technologies have allowed significant reductions in the material budget with respect to hybrid pixel detectors, while improving the signal-to-noise ratio and the position resolution that is achievable with CMOS sensors.

However, the simulation of such devices remains challenging due to the complex field configuration in the sensor.
Advanced simulation tools are required to understand and model the performance of detectors built in these technologies and to optimize the design of future prototypes.

This paper presents a simulation performed with a combination of commonly used tools employed in silicon detector simulation.
The \apsq framework~\cite{apsq} is used to combine TCAD-simulated electric fields with a Geant4~\cite{geant4,geant4-2,geant4-3} simulation of the particle interaction with matter, to investigate the behavior of high-resistivity CMOS detectors and to compare the predicted performance with measurements recorded in a particle beam.

This allows direct access to detector performance parameters such as spatial resolution and detection efficiency by taking into account the stochastic nature of the initial energy deposition.
While many of these properties could also be investigated by advanced TCAD transient simulations, this approach is not practical owing to the high computing time for a single event and the high-statistics samples required to evaluate the effects related to the strong variation of the electric field in three dimensions.

Instead, a simplified charge transport algorithm is used, taking as an input the electrostatic field map calculated by the TCAD simulation of the complex field configuration within the sensor.
The algorithm takes into account effects like Landau fluctuations in the energy deposition and the production of secondary particles such as delta rays.
With event simulation rates of several tens of Hertz, this allows for the generation of high-statistics samples necessary for detailed studies of the detector behavior.

The paper is structured as follows.
Section~\ref{sec:technology} provides a brief overview of the CMOS process under investigation, while the detector properties and the simulated setup are introduced in Section~\ref{sec:detector_design}.
The simulation is described in detail in Section~\ref{sec:simulation}, while Section~\ref{sec:reconstruction} introduces the reconstruction of physical properties from the detector response.
The sensitivity of the simulation to a range of parameters is examined in Section~\ref{sec:error}.
The simulation is validated using data recorded in test-beam measurements in Section~\ref{sec:validation}, while performance quantities are derived in Section~\ref{sec:performance} and compared with the values obtained from data.
Finally, Section~\ref{sec:summary} summarizes the results and provides an outlook for future investigations of this technology.


\section{The High-Resistivity CMOS Process}
\label{sec:technology}
\begin{figure}[tbp]
	\centering
	\includegraphics[width=1\columnwidth]{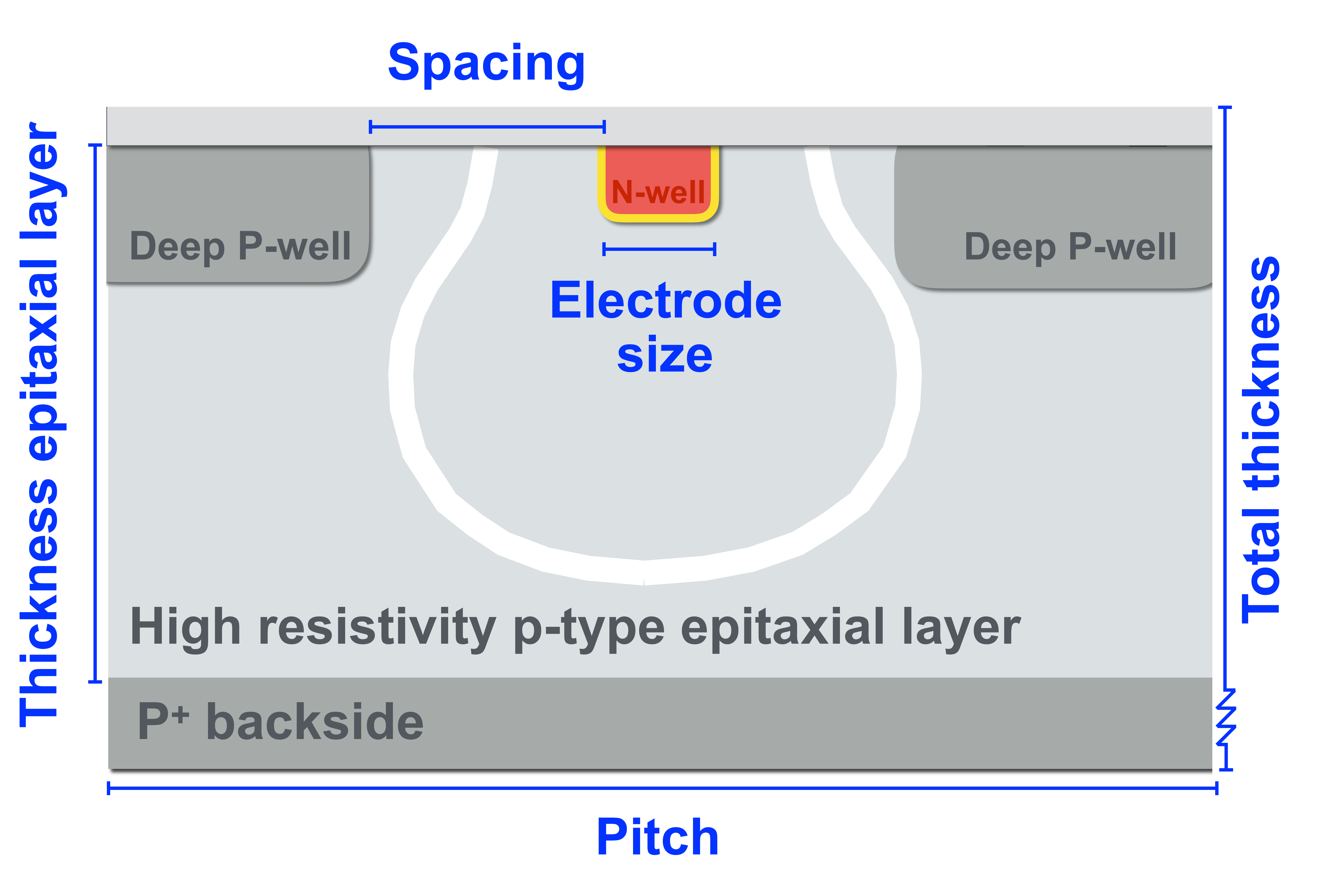}
	\caption{Schematic cross section of a single pixel cell in the CMOS process under investigation. The elements shown are not to scale. Modified from~\cite{DANNHEIM2019187}.}
	\label{fig:schematic_crosssection}
\end{figure}

Monolithic CMOS technologies incorporating the readout electronics in the sensor are attractive candidates for new detector designs to simplify the production and to benefit from a reduction of the material budget.
By integrating the CMOS logic in doped wells separated from the inversely doped signal collection electrode, the size of the latter can be minimized as illustrated in Figure~\ref{fig:schematic_crosssection}.
The small collection electrode design allows the sensor capacitance to be reduced down to the order of \si{\femto \farad}, enabling detector designs with low noise and detection thresholds, low analog power consumption, and large signal-to-noise ratio (SNR)~\cite{Snoeys}.

Implemented in a standard silicon substrate, only a small depleted region evolves around the \emph{pn}-junction surrounding the collection electrode when applying a bias voltage between the doped wells and the backside of the sensor.
The applicable bias voltage is limited to \SI{-6}{V} by the process-specific breakdown voltage of the NMOS transistors~\cite{thesis-jacobus}.
In order to achieve a sizable depletion volume around the collection electrode, an epitaxial layer with high resistivity silicon can be used.

The size of the depleted region forming in this epitaxial layer is restricted to the area around the collection electrode and, without additional modifications of the process, no full depletion of the sensor volume is achieved.
In the CMOS process under investigation, the depleted region has the shape of a bubble as indicated by the white line in Figure~\ref{fig:schematic_crosssection}, resulting in contributions to the overall detector response from both drift and diffusion of charge carriers.
In addition, signal contributions are expected from charge carriers that are created in the highly \emph{p}-doped backside substrate and subsequently diffuse into the epitaxial layer.


\section{Detector Design under Investigation}
\label{sec:detector_design}
The \emph{Investigator} test-chip is an analog test chip that has been developed within the ALICE ITS upgrade~\cite{alice-its}.
It has been investigated by the CLICdp collaboration to evaluate this technology in terms of sensor performance focussing on precise measurements of spatial resolution and detection efficiency~\cite{DANNHEIM2019187,thesis-magdalena}.
The digitization of signals is performed off-chip in the data acquisition system using one \SI{65}{\MHz} sampling analog-to-digital converter (ADC) per channel which records the full waveform of all detector channels, once a configurable triggering threshold has been exceeded in any of them~\cite{thesis-krzysztof}.
It should be noted that the threshold values for data quoted below represent the offline analysis thresholds applied in addition to the triggering threshold of about \SI{120}{e}.

The chip has a total thickness of \SI{100}{\micro \meter}.
The upper \SI{25}{\micro \meter} of the sensor, below the implants, consist of the epitaxially grown silicon with a resisitiviy of $1 - \SI{8}{\kilo \ohm \centi \meter}$ in which the depleted region forms, while the additional \SI{75}{\micro \meter} represent the undepleted low-resistivity silicon substrate~\cite{thesis-jacobus}.

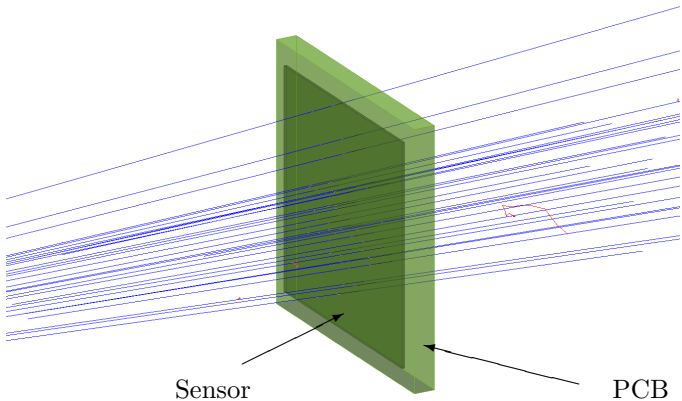
\begin{figure}[tbp]
  \centering
  \begin{overpic}[width=\columnwidth]{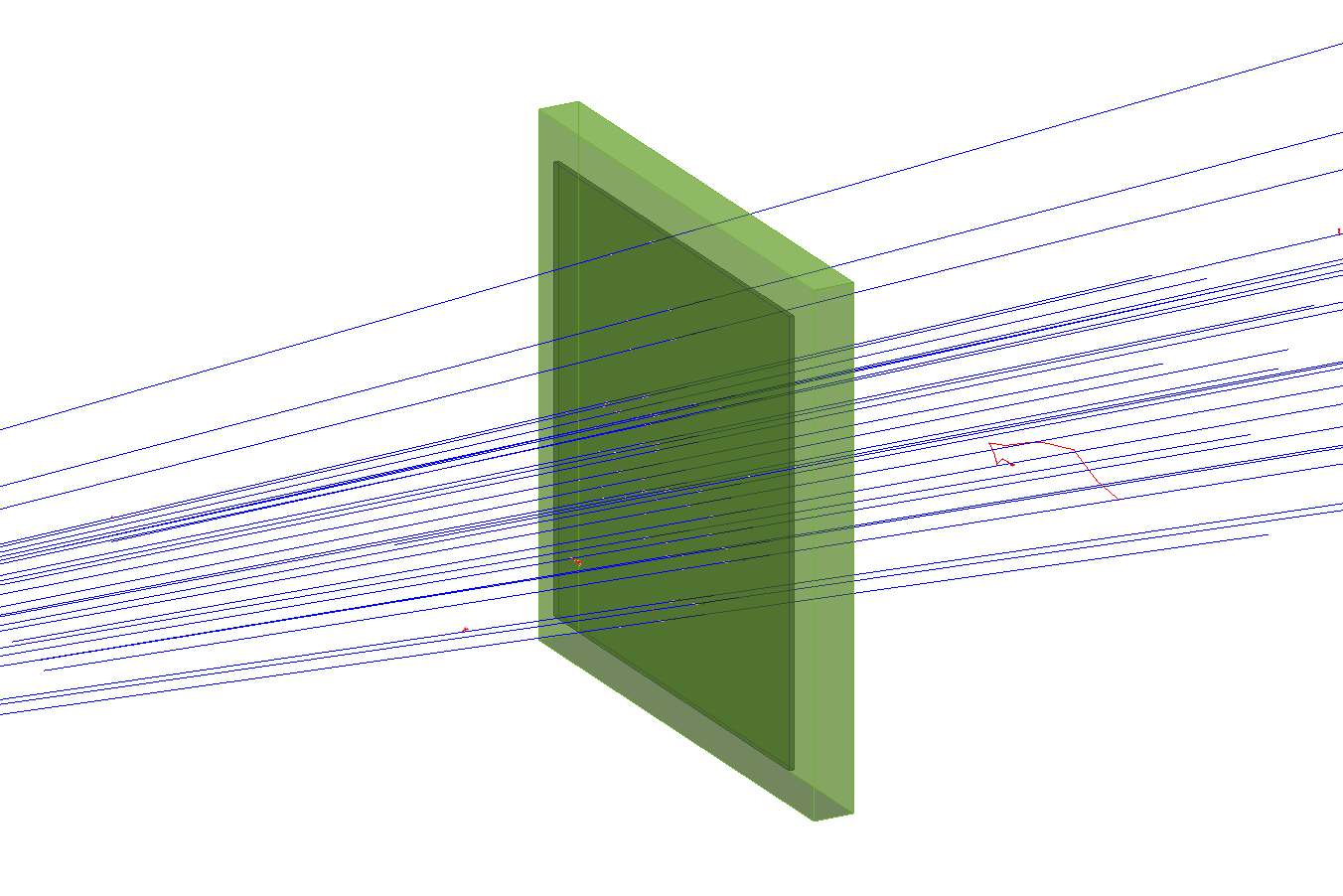}
      \put (25,5) {Sensor}
      \put (35,10) {\vector(2,1){15}}
      \put (90,5) {PCB}
      \put (85,7) {\vector(-4,1){23}}
  \end{overpic}
  \caption{Visualization of the simulated detector setup consisting of the CMOS sensor on a printed circuit board for support. The detector is oriented perpendicular to the beam incident from the left. The colored lines represent the primary and secondary particles propagated through the setup.}
  \label{fig:visualization}
\end{figure}

While the actual detector contains several sub-matrices with \num{8 x 8} active pixels each, with different layouts such as altered collection electrode size, only one matrix has been simulated and is compared to data.
The pixel cells of the chosen matrix have a pitch of \SI{28 x 28}{\um} and feature the following geometrical parameters:
the distance between the \emph{p}-wells and the collection electrode is \SI{3}{\micro \meter} and an octagonal collection electrode with a size of \SI{2x2}{\micro \meter} is placed in the center of the pixel cell.
A bias voltage of \SI{-6}{V} is applied to the \emph{p}-wells and a positive voltage of \SI{+0.8}{V} is applied to the collection electrode itself.
The simulated detector is placed on a printed circuit board (PCB) as visualized in Figure~\ref{fig:visualization}.


\section{Simulation Flow}
\label{sec:simulation}
In the following section, the simulation of the detector in the \apsq framework is described.
In order to avoid simulating a full beam telescope setup and performing a track reconstruction, the capabilities of the framework to record the Monte Carlo truth information about primary and secondary particles are exploited.

Consequently, only a single CMOS detector and the source of ionizing particles are simulated as shown in Figure~\ref{fig:visualization}.
The figure depicts the overlay of many events, as only a single primary particle is simulated in each event.

The following sections describe the individual steps of the simulation in detail, providing information on the configuration of the respective \apsq modules where applicable and relevant.

\subsection{Electrostatic Field Modeling with TCAD}

The electrostatic field in the epitaxial layer of the sensor is modeled using a three-dimensional TCAD simulation.
The doping profile is taken from~\cite{thesis-magdalena, pixel-proceedings} and resembles the technology described in Section~\ref{sec:technology}, with the detector geometry introduced in Section~\ref{sec:detector_design}.
The simulation comprises a single pixel cell, and periodic boundary conditions allow the field to be replicated over the entire sensor.

\begin{figure}[tbp]
	\centering
	\includegraphics[width=1\columnwidth]{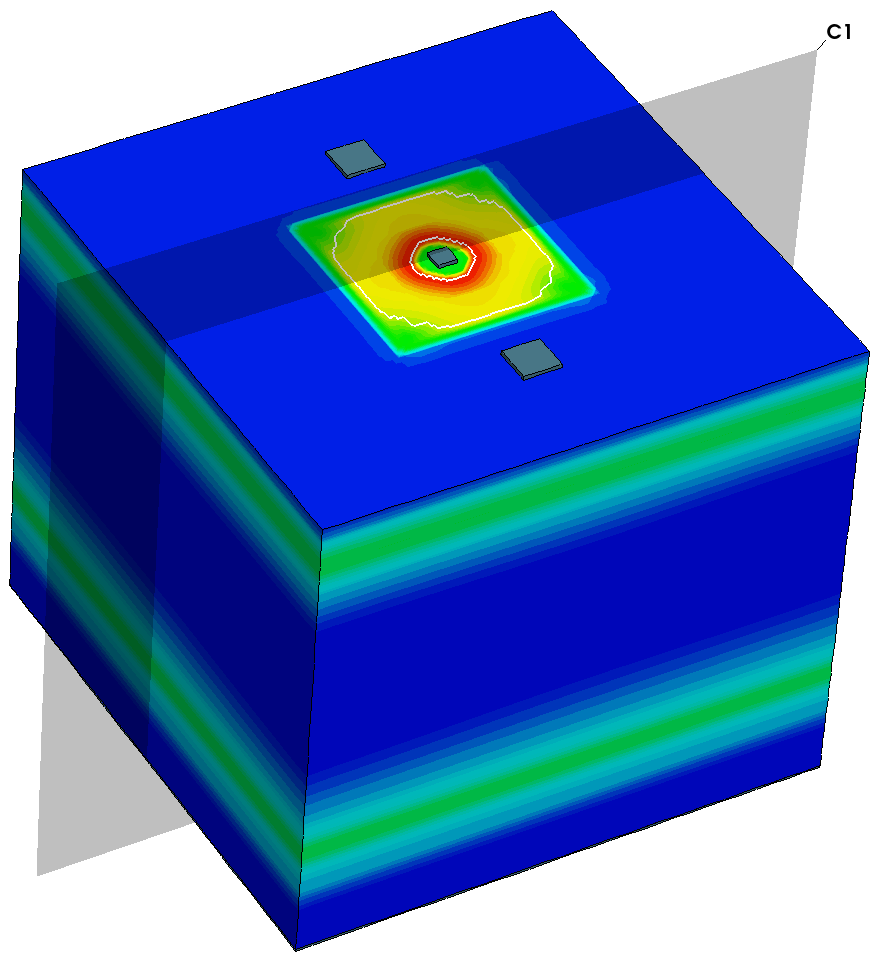}
	\caption{Magnitude of the electric field inside the pixel cell, simulated using TCAD. The visualization only shows the upper \SI{25}{\micro \meter} of the sensor with the epitaxial layer. The gray structures represent metal contacts used as terminals for the biasing voltages. The plane C1 indicated in gray corresponds to the cut presented in Figure~\ref{fig:efieldlines} (color online).}
	\label{fig:efield}
\end{figure}

Figure~\ref{fig:efield} shows a visualization of the magnitude of the electric field in the three-dimensional pixel cell, with the corresponding voltages applied to the terminals via metal contacts indicated as gray structures.
A low electric field is present in the \emph{p}-well rings as indicated by the blue region on the surface of the simulated pixel cell.
The center of the \emph{p}-well rings is fitted with a squared opening that contains the collection electrode with a high-field region evolving around it.

\begin{figure}[tbp]
	\centering
	\includegraphics[width=\columnwidth]{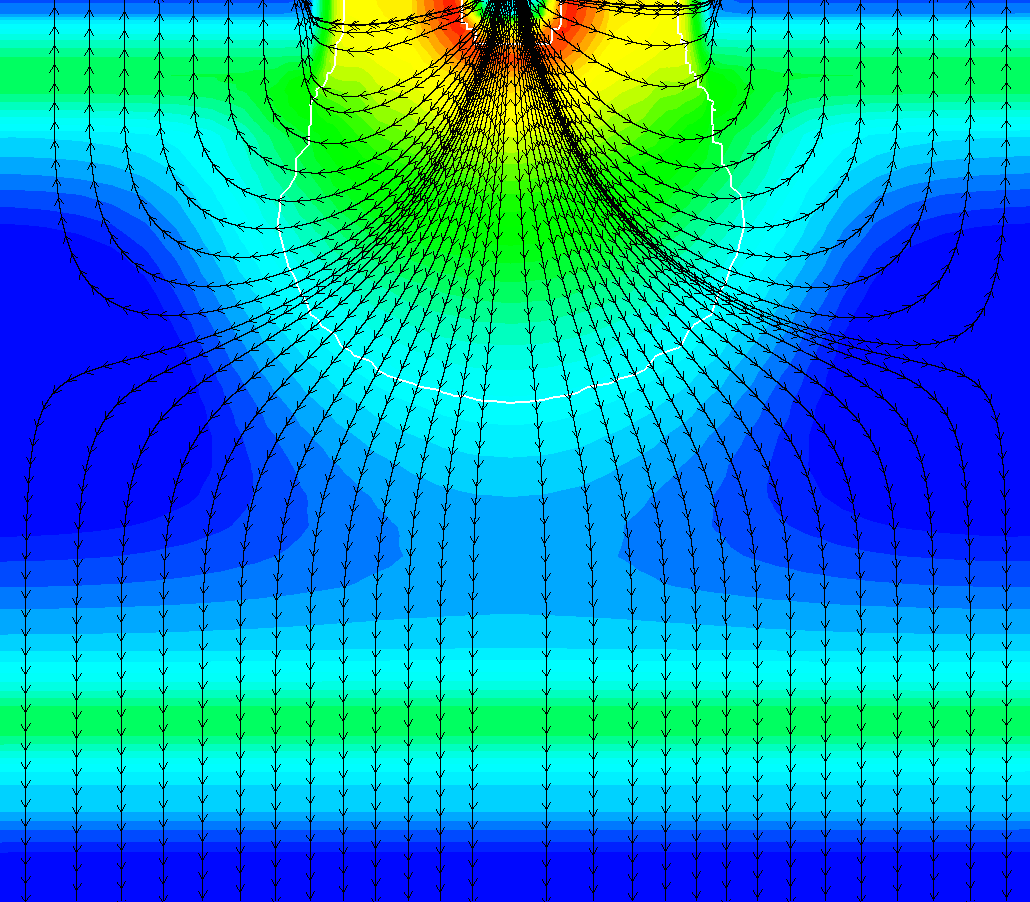}
	\caption{Magnitude of the electric field and field lines for a cut through the 3D TCAD simulation. The plot only depicts the upper \SI{25}{\micro \meter} of the sensor with the epitaxial layer, while the undepleted substrate region is omitted (color online).}
	\label{fig:efieldlines}
\end{figure}

The strong inhomogeneities of the electric field in different regions of the pixel cell are best observed in a cut through the collection electrode, perpendicular to the sensor surface, as depicted in Figure~\ref{fig:efieldlines}.
The high electric field strength close to the \emph{pn}-junction around the collection electrode decreases rapidly towards the sensor backside and the pixel corners.
The white line indicates the depleted volume of the pixel cell.
The electric field lines, indicated as black arrows, provide a first insight into the complexity of the field configuration in the sensor and the drift effects induced by this strong non-linearity.
The low electric field regions in the pixel corners result in a slower charge carrier drift and an increased impact of diffusion, leading to an enhanced charge sharing which improves the position resolution without the need to reduce the pixel pitch.
In the low-resistivity substrate, recombination of charge carriers is a relevant process owing to the higher doping concentration.


The electrostatic field obtained from TCAD is converted to a regularly spaced mesh using the \emph{Mesh Converter} tool provided with the \apsq framework.
This conversion speeds up the look-up of field values during the simulation by several orders of magnitude since all necessary interpolations are already performed offline prior to the simulation.
A regular mesh granularity of \SI{0.1}{\um} is chosen to correctly replicate the field in the high-density mesh regions of the TCAD simulation close to the implant.

It has been verified that the selected granularity correctly replicates the TCAD simulation by comparing the two fields.
Using an even finer granularity has not shown any significant improvement on the simulation results.
Loading highly granular electrostatic fields in \apsq does not impact the performance of the simulation, but only the memory footprint of the program during execution.

\subsection{Energy Deposition with Geant4}

\apsq provides the \emph{DepositionGeant4} module, an interface to Geant4~\cite{geant4,geant4-2,geant4-3} which facilitates the simulation of energy deposition in the sensor.
A \SI{120}{\GeV} beam of \Ppiplus incident on the pixel detector is simulated, replicating the beam conditions of the test-beam measurements.
The beam points along the positive \emph{z}-axis, perpendicular to the \emph{xy}-plane of the detector.
The cross section of the beam is chosen to be significantly smaller than the detector surface to suppress effects stemming from the altered charge sharing behavior at the sensor edge.
The energy deposited in the sensor by Geant4 is translated into charge carriers with a conversion factor of \SI{3.64}{\eV} per electron-hole pair~\cite{chargecreation}.

The framework also stores the Monte Carlo truth information including secondary particles such as delta rays and their relation to the primary particles.
This information can be exploited to establish a link between the incident particles and the electron-hole pairs created in the detector.

The simulation is performed with the Photo-Absorption Ionization model (PAI)~\cite{pai} to improve the description of energy deposition in thin sensors.
This is of importance in spite of the total sensor thickness of \SI{100}{\um}, since a majority of the charge carriers forming the final signal will originate from the \SI{25}{\um} epitaxial layer.

\begin{listing}
\begin{minted}[frame=single,framesep=3pt,breaklines=true,tabsize=2]{ini}
[DepositionGeant4]
physics_list        = "FTFP_BERT_EMY"
enable_pai          = true
particle_type       = "Pi+"
source_type         = "beam"
source_energy       = 120GeV
source_position     = 0um 0um -200um
beam_size           = 0.5mm
beam_direction      = 0 0 1
number_of_particles = 1
max_step_length     = 1.0um
\end{minted}
\caption{Configuration section for the \emph{DepositionGeant4} module setting up the particle source in Geant4 for the initial energy deposition in the sensor.}
\label{lst:depositiongeant4}
\end{listing}

The module configuration used for the \apsq framework is provided in Listing~\ref{lst:depositiongeant4}.

\subsection{Charge Carrier Transport}

The signal formation is simulated using a simplified model for charge carrier transport based on the notion of \emph{collected charges}.
The electron-hole pairs created by the incident particle are propagated along the electric field lines through the sensor volume using an adaptive fourth-order Runge-Kutta-Fehlberg (RKF) method~\cite{fehlberg} and a mobility parametrization which depends on the electric field vector~\cite{JACOBONI197777}.
The RKF method adapts the simulated time step depending on the position uncertainty derived from a fifth-order error estimation; the allowed range for time steps was set to $\SI{0.5}{\pico \second} \leq \Delta t \leq \SI{0.5}{\nano \second}$.

While this model is not expected to reproduce a realistic time dependence of the signal, the final state of charge collected at the sensor implants is equivalent to the integrated induced current over the respective drift time.
This approximation is valid since the Shockley-Ramo weighting field~\cite{shockley,ramo} is negligible in most of the sensor volume owing to the small ratio between signal collection electrode size and sensor thickness.

In the upper \SI{25}{\micro m} of the sensor the charge carrier motion is a superposition of drift and diffusion, while in the lower \SI{75}{\micro m} the charge carriers are only subject to random diffusion as the electric field is negligible.

The propagation algorithm is halted after \SI{22.5}{\nano \second}, the so-called \textit{integration time}, and all charge carriers within a volume of $3 \times 3 \times \SI{2}{\micro \meter^3}$ around each of the signal collection electrodes are attributed to the respective pixel signal.
The volume has been chosen to cover the electrode implant itself as well as an additional volume accounting for the uncertainty in the final position of the transported charge carriers.
The integration time has been chosen such that the simulation produces clusters with the same most probable value (MPV) for the cluster charge as obtained from data.
This aims to emulate the physical process of charge carrier recombination in the silicon substrate, which might be modeled directly in future simulations as briefly discussed in Section~\ref{sec:summary}.
The systematic uncertainty introduced by this approach is discussed in Section~\ref{sec:error}.

Charge carriers are transported in groups of five instead of individually to speed up the simulation process.
The group size has been chosen such that an adequate number of transport steps is retained with the expected MPV for the signal of around \SI{1.5}{\kilo e}.
It has been verified that this simplification does not affect the simulation result as further elaborated in Section~\ref{sec:error}.

\begin{figure*}[tbp]
 \centering
 \includegraphics[width=\textwidth]{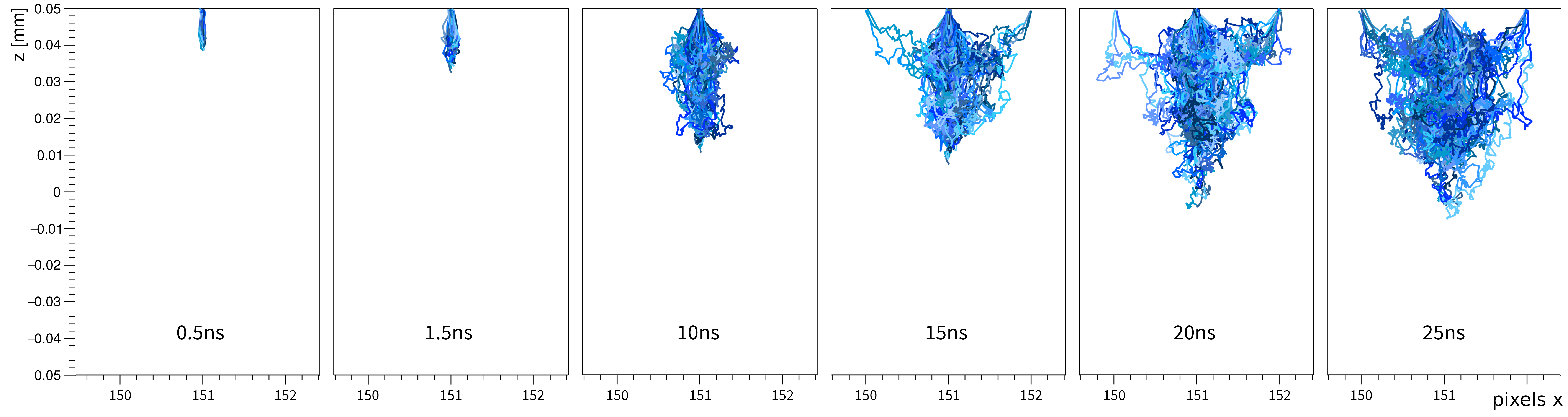}
 \caption{Visualization of the time evolution of the collected charge. Shown are snapshots at different times after the initial energy deposition with each line representing the drift and diffusion motion of a group of charge carriers. Only charge carrier groups which have reached the implant are drawn, all other charge carriers are omitted. The ionizing particle traverses the sensor along the $z$-axis in the center of a pixel cell, each plot represents three adjacent pixels.}
 \label{fig:lineplot}
\end{figure*}

Figure~\ref{fig:lineplot} visualizes this transport model and shows the collection of charge carriers at the electrodes of the sensor.
In this representation, only electrons that have reached a sensor implant within the integration time are shown.
Electrons that are still in motion as well as holes are suppressed.
The motion of each group of charge carriers is represented by one line and is shown at different integration times after the initial energy deposition.
Here, the incident particle traversed the detector along the $z$-axis through the center of one pixel cell.

After the first few hundred picoseconds, only charge carriers in the vicinity of the electrode are collected.
The straight lines indicate that their motion is dominated by drift.
With increasing integration time, the motion patterns of further groups of charge carriers arriving at the implant exhibit a strong contribution from diffusion as indicated by the numerous kinks in the respective paths.
After about \SI{15}{\ns}, lateral motion enables some charge carriers to be collected in the two adjacent pixel cells.

The line graphs also allow visual distinction between the substrate and the epitaxially grown high-resistivity layer, which ends about \SI{25}{\um} from the top of the sensor.
A faster drift motion can be observed in the high-field region close to the backside of the epitaxial layer as straight lines; the contribution from substrate charge carriers diffusing into the epitaxial layer starts only after approximately \SI{10}{\ns}.

\begin{figure}[tbp]
 \centering
 \includegraphics[width=\columnwidth]{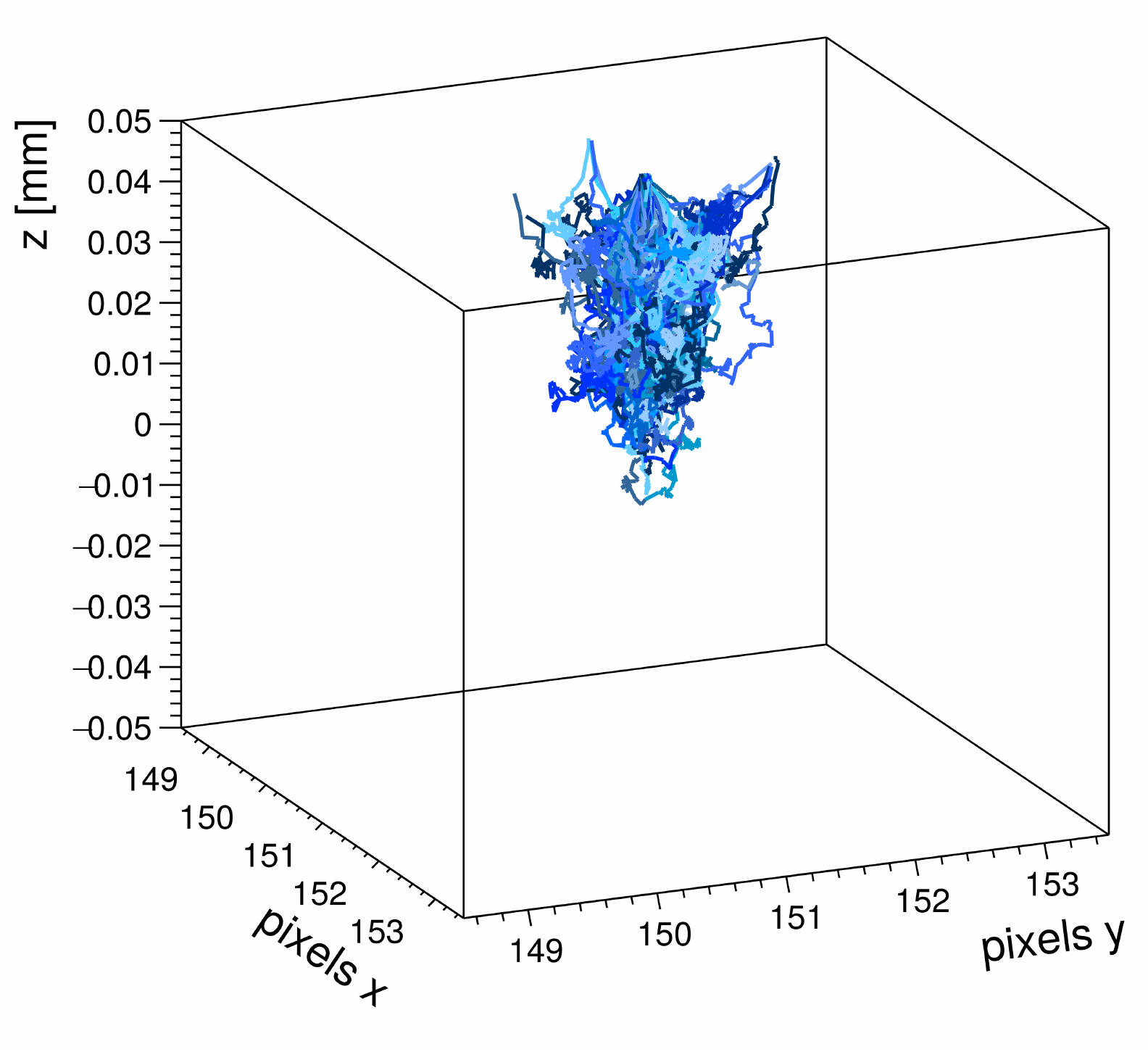}
 \caption{Three-dimensional visualization of the charge carrier motion, corresponding to the \SI{20}{\nano \second} snapshot shown as projection in Figure~\ref{fig:lineplot}.}
 \label{fig:lineplot3d}
\end{figure}

In Figure~\ref{fig:lineplot3d}, a three-dimensional representation of the line plot at \SI{20}{\ns} is presented.
The lines end at five different points, each representing a different collection electrode.

\begin{listing}
\begin{minted}[frame=single,framesep=3pt,breaklines=true,tabsize=2]{ini}
[GenericPropagation]
temperature      = 293K
charge_per_step  = 5
timestep_min     = 0.5ps
timestep_max     = 0.5ns
integration_time = 20ns
\end{minted}
\caption{Configuration section for the \emph{GenericPropagation} module used to simulate the charge transport.}
\label{lst:genericpropagation}
\end{listing}

The configuration provided in Listing~\ref{lst:genericpropagation} has been used for the charge carrier transport.
Settings for creating line graphs of the charge carrier motion can be found in the \apsq user manual available from the project website~\cite{apsq-website}.

\subsection{Digitization of Signals}

To simulate the response of the readout electronics, the charge carriers accumulated in the region around the signal collection electrode during the integration time are transformed into a digital signal.
While the detector under investigation uses off-chip ADCs for the signal digitization as described in Section~\ref{sec:detector_design}, the simulation aims to simulate an on-chip per-pixel threshold using the \emph{DefaultDigitizer} module of \apsq.
Equivalent noise values have been used where applicable, as discussed below.

\begin{listing}
	\begin{minted}[frame=single,framesep=3pt,breaklines=true,tabsize=2]{ini}
[DefaultDigitizer]
electronics_noise  = 10e
threshold          = 40e
threshold_smearing = 5e
\end{minted}
\caption{Configuration section used for the \emph{DefaultDigitizer} module of the simulation.}
\label{lst:defaultdigitizer}
\end{listing}

An additional signal contribution, randomly drawn from a Gaussian distribution with a width of \SI{10}{e} and a mean of \SI{0}{e} is added to the signal to account for electronics noise present during digitization.
The applied threshold is varied between \SI{40}{e} and \SI{700}{e}, and a threshold dispersion, sampled from a Gaussian distribution with a width of \SI{5}{e} and a mean of \SI{0}{e}, is added.
For simplicity, the threshold dispersion is not a fixed offset calculated per-pixel, but randomly chosen per pixel hit.
The setup of the module is summarized in Listing~\ref{lst:defaultdigitizer}.

\subsection{Data Processing and Storage}

The simulation results are stored in ROOT~\cite{root} \texttt{trees} using the \emph{ROOTObjectWriter} module.
In order to speed up the process, the simulation is performed in two separate steps.
In the first step, the energy deposition, charge carrier transport and summing of charge at the collection electrodes is performed.
The result of this step is stored to disk.

In a second step, the \emph{ROOTObjectReader} is used to read the information from the respective file and the final digitization step is performed.
This allows to re-run this final section of the simulation on the full set of Monte Carlo events with different settings applied without the need to recompute the drift motions.
A full threshold scan, performed on the same set of initial simulation events, thus only takes a couple of minutes instead of several hours required to create the initial data set.
Since the threshold scan performed on the test-beam data has also been performed offline on the same data set~\cite{thesis-magdalena}, this is an adequate simplification of the simulation.

The central simulation data set comprises about 2.5 million primary events which have been reprocessed for every threshold setting.
In addition, several smaller data sets with different integration times have been produced in order to optimize agreement with data as discussed in Section~\ref{sec:error}.


\section{Reconstruction and Analysis}
\label{sec:reconstruction}
In the following, the reconstruction and analysis of the Monte Carlo events are discussed.
The simulation was set up using known, independent parameters of the measurement setup, such as track resolution or charge threshold.
Only the cluster charge MPV was used as direct observable provided by the detector to tune the simulation.
All parameters were fixed before comparison with data for observables used to quantify the performance, such as cluster size, position resolution and efficiency.
This blinded approach avoids drawing premature conclusions from the figures of merit and thus influencing the parameter optimization.
Using only the MPV of the cluster charge for calibrating the simulation minimizes the correlation between simulation and data, and maximizes the prediction power of the simulation.

\subsection{Reference tracks}

The Monte Carlo truth information provided by the \apsq framework is used as reference track information.
All registered particles in the sensor are filtered and only primary particles entering the sensor from the outside, i.e.\ those without a parent particle, are selected for further analysis.
This set of particles represents external tracks, and their position in the mid-plane of the sensor is calculated by linearly interpolating their entry and exit points registered by the framework.
This position is then convolved with the track resolution at the device under test (DUT) of \SI{2.0}{\um}, in accordance with the value obtained for the beam telescope used for the acquisition of the test-beam data~\cite{AlipourTehrani}.

\subsection{Clustering}

The pixel hits registered by the simulated detector are grouped into clusters by starting with a seed pixel and adding all directly adjacent pixel hits to the cluster until no additional neighbors are found.
This already allows basic properties of the simulation to be compared with data, namely cluster size as well as the shape of the cluster charge distribution.

The total cluster charge is given by the sum of the individual pixel charges of the cluster.
Its comparison with data allows the required integration time in the simplified simulation model to be adjusted to achieve the same integrated charge as seen in data.
This procedure is described in detail in Section~\ref{sec:error}.

The cluster size is defined as the total number of pixels contained in the respective cluster.
It has a strong dependence on the drift and diffusion of the charge carriers in the sensors and is the primary measure for charge sharing between pixel cells.
It thus allows evaluation of the performance of the simulation, e.g.\ how well the electric field is modeled.

\subsection{Reconstruction of the Cluster Position}

For assessing the performance of the detector, a particle incidence position has to be extracted from the cluster information available.
To replicate the analysis performed for the test-beam data, the charge-weighted center-of-gravity position of the cluster is corrected for non-linear charge sharing by an $\eta$~algorithm~\cite{Belau1983253}.

\begin{figure}[tbp]
	\centering
	\includegraphics[width=\columnwidth]{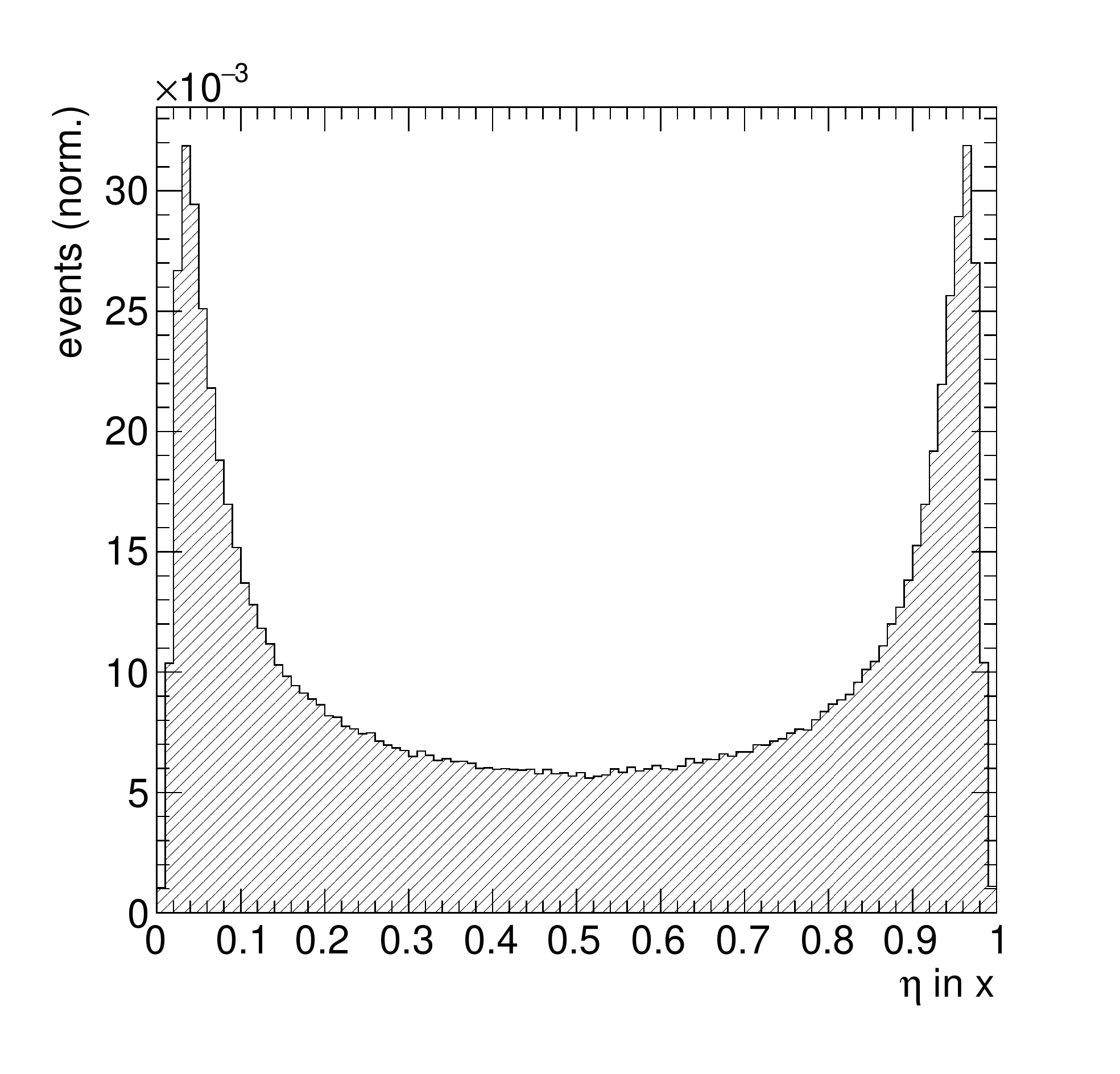}
	\caption{$\eta$-distribution along the $x$ axis, derived from simulation at a threshold of \SI{40}{e}.}
	\label{fig:eta_x}
\end{figure}

Since the $\eta$~distribution represents the charge sharing between two pixels only, for each cluster the two pixels with the highest charge, $Q_1$ and $Q_2$, are chosen to construct the $\eta$ variable independently in $x$ and $y$:
\begin{align*}
\eta_{k} = \frac{\sum_{i} k_{i} \cdot Q_i}{\sum_{i} Q_i} \qquad k = \{x, y\} \quad i = \{1, 2\}
\end{align*}
where $k_{i}$ is the relative position between the two pixel centers.
An example of the $\eta$ distribution in $x$ is depicted in Figure~\ref{fig:eta_x} for a pixel charge threshold of \SI{40}{e}.


\section{Systematic Uncertainties}
\label{sec:error}
The sensitivity of the simulation to different input parameters has been examined by varying the values within their respective uncertainties, if known, or within a reasonable range otherwise.
The impact on the reconstructed observables was investigated.
While some parameters exhibit little or no effect on the simulation results, others have a strong influence on the outcome.

\subsection{Free parameters}

For the initial deposition of energy in the sensor, the influence of the maximum allowed step length of tracking primary and secondary particles through the sensor material has been evaluated by varying the respective value between \SI{0.1}{\micro \meter} and \SI{5}{\micro \meter}, and no significant difference was observed.
Since large parts of the sensor volume are undepleted, a strong impact of diffusion is expected which smears the initial position of the charge carriers.

The charge carrier transport is mainly dominated by the precision of the numeric integration and its granularity.
The number of charge carriers transported as group has been varied from a single charge carrier up to ten per group in order to study possible effects on the distribution at the implants.
The effect on the reconstruction observables is found to be negligible.

\subsection{Parameters constrained by measurements}

The behavior of the sensor has been shown to be very sensitive to the simulated physical properties of the CMOS sensor, i.e.\ the thickness of the epitaxial layer as well as the modeled electric field.
Even small changes in the sensor design, such as a more simplistic approximation of the implant doping profiles in the TCAD design cause large changes in the resulting cluster size distributions and position resolution.
It is therefore of paramount importance to model the sensor as precisely as possible and to constrain the different parameters in TCAD by additional measurements~\cite{tj-modified}.
The low-field regions found in the corners of the pixel cell visible in Figures~\ref{fig:efield} and~\ref{fig:efieldlines} are strongly influenced by these modifications, and their contribution to the detector signal changes accordingly.

\begin{figure}[tbp]
 \centering
 \includegraphics[width=\columnwidth]{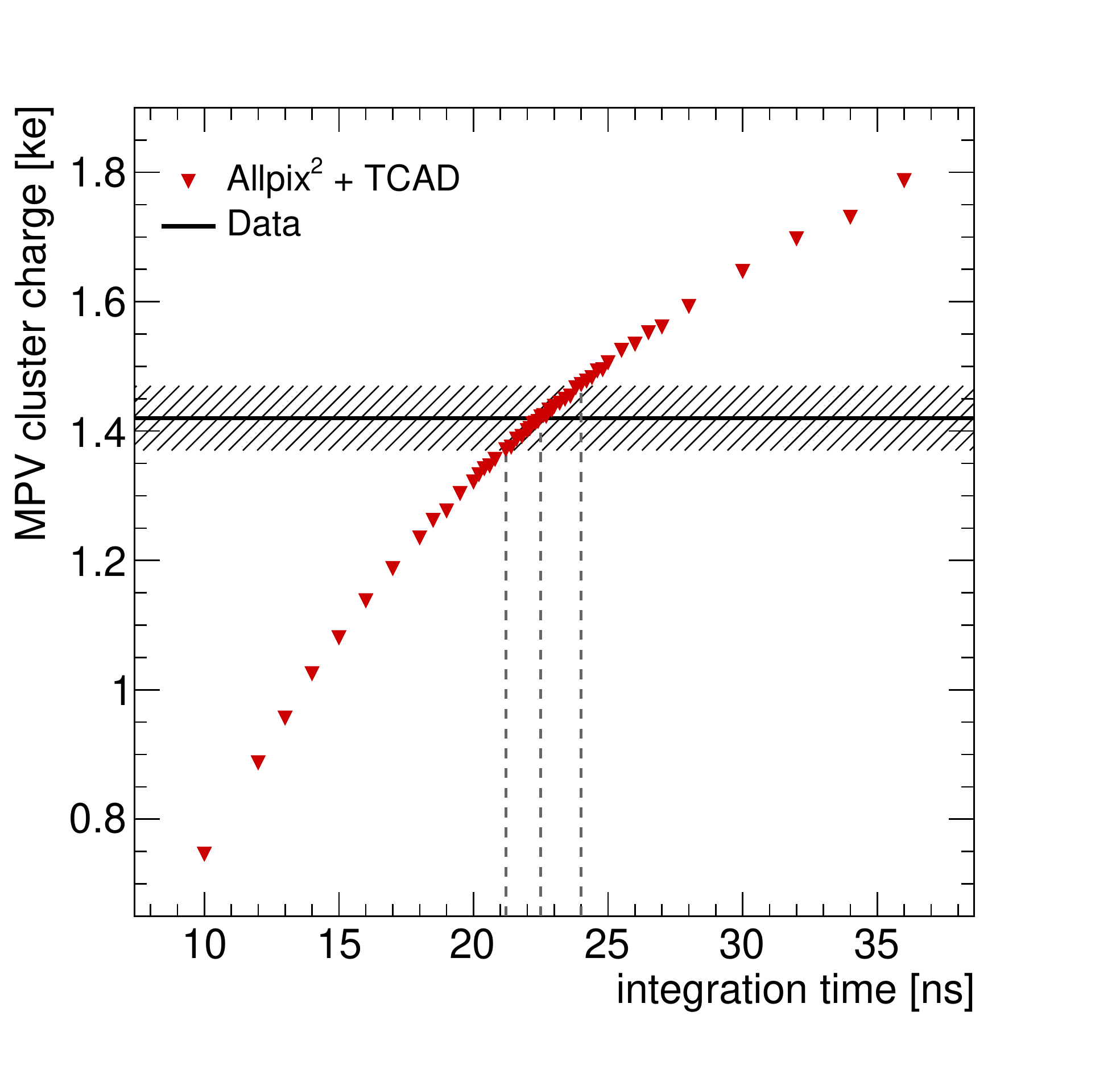}
 \caption{Most probable cluster charge as a function of the simulated integration time. The black horizontal line represents the value obtained from data, the hatched band corresponds to the assumed systematic uncertainty of the charge calibration.}
 \label{fig:integrationscan}
\end{figure}

The integration time currently used to stop the transport of charge carriers is also linked to the sensor design, since it is used to emulate an effective recombination of charge carriers.
Their lifetime in the different regions of the sensor is dominated by the respective doping concentration, and potentially affected by the silicon wafer production process.
Since this was not modeled in detail for this simulation, the integration time was chosen such that the MPV of the cluster charge matched the value obtained from data as discussed in Section~\ref{sec:simulation}.
The corresponding uncertainty on the charge calibration of the reference data has therefore to be taken into account as systematic uncertainty of the simulation by comparing the cluster charge MPV for different integration times to the value obtained from data as shown in Figure~\ref{fig:integrationscan}.
Here, the hatched band represents an assumed uncertainty of \SI{+-50}{e} on the charge calibration of data~\cite{thesis-jacobus, thesis-magdalena}.
This translates to an uncertainty on the integration time of $22.5^{+1.5}_{-1.3}\,\textrm{ns}$, which is propagated as systematic uncertainty to the results presented in this paper.

It has been observed that the overall agreement between data and simulation seems to improve for lower integration times, which might indicate either an offset in the absolute charge calibration of data or an insufficient modeling of the signal formation processes in silicon.

Also the charge threshold applied to the individual pixels has a strong impact on both the cluster size and the intrinsic resolution, with decreasing influence towards higher thresholds.
At a threshold of \SI{40}{e}, a change of as little as \SI{\pm5}{e} visibly alters the observables.
Since the absolute charge calibration and the threshold value in electrons are fully correlated, the uncertainty on the applied threshold has been taken into account by varying the two parameters simultaneously and by calculating the total uncertainty arising from the variations.

A variation of the threshold dispersion and electronics noise of up to \SI{10}{e} at a threshold of \SI{40}{e} yielded no observable effect.
The values for noise and threshold dispersion have been estimated from the evaluation of the full waveform in data~\cite{thesis-magdalena}.

The residual width and the final intrinsic resolution depend on the resolution of the reference tracks at the position of the detector under investigation.
This resolution has been determined for the test-beam data used, and a variation of \SI{\pm0.2}{\micro \meter} around this value shifts the obtained resolution accordingly.
This strong influence arises from the fact that the two values are of similar size.

In summary, while the free parameters of the simulation have little to no influence on the final result when varied within a reasonable range, several parameters show a high sensitivity but are constrained by measurements.


\section{Validation With Test-Beam Data}
\label{sec:validation}
The simulation is compared to data recorded with the \emph{Investigator} chip, described in Section~\ref{sec:detector_design}, at the CERN SPS accelerator with a \SI{120}{GeV} \Ppiplus beam.
A total of 25660~tracks through the region of interest have been recorded, mainly limited by the very small active area of the DUT and the dead time of the data acquisition system used.
More details about the test-beam setup, data samples and the analysis of data used for comparison in this paper can be found in~\cite{DANNHEIM2019187,thesis-magdalena}.

\subsection{Cluster Charge}

\begin{figure}[tbp]
	\centering
	\includegraphics[width=\columnwidth]{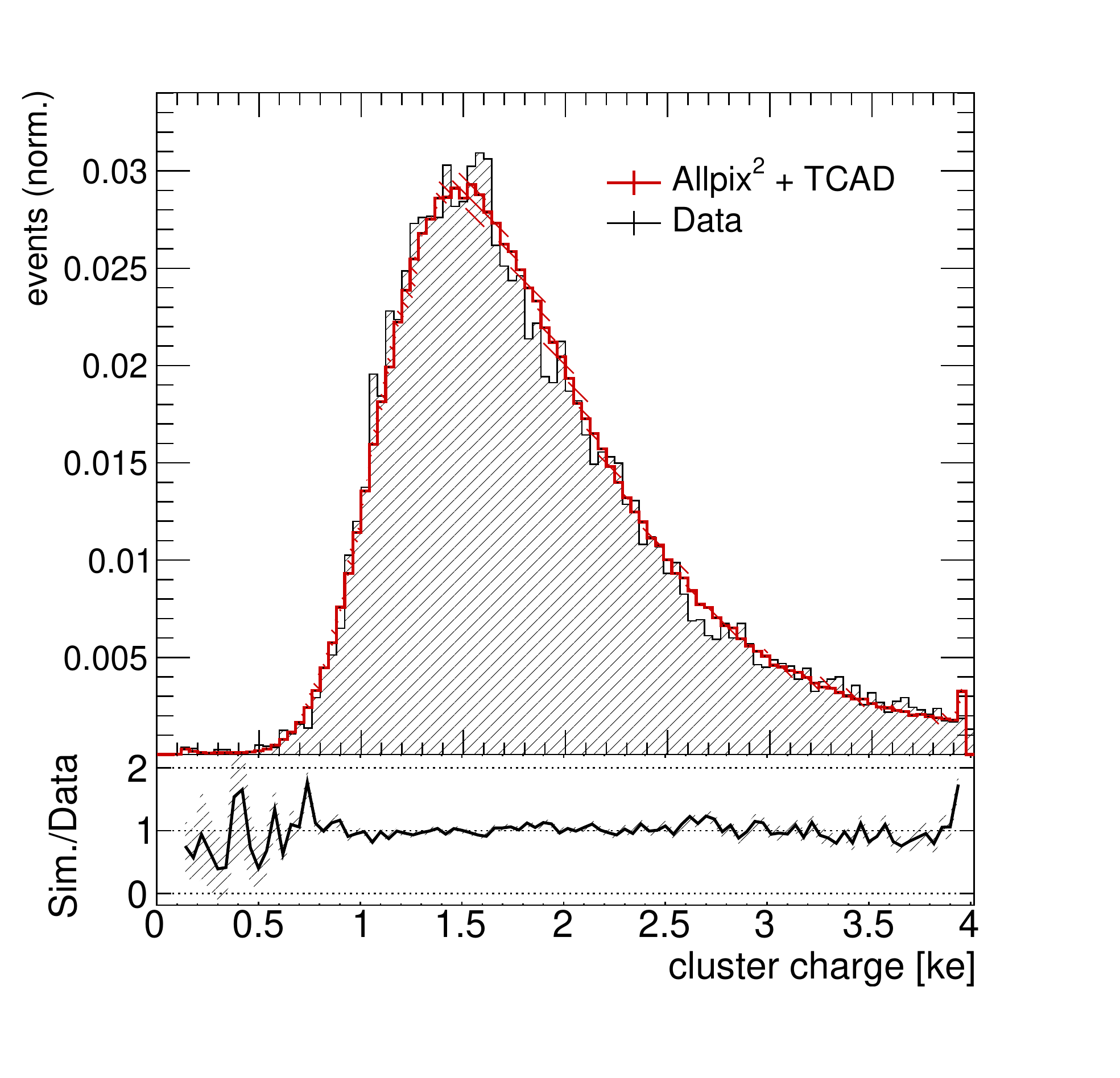}
	\caption{Cluster charge distributions at a pixel threshold of \SI{120}{e} for simulation and experiment.
	The distributions resemble the expected Landau-Gauss distribution. The hatched band represents the total uncertainty.}
	\label{fig:cluster_charge_dis}
\end{figure}

The cluster charge distributions for both simulation and data at a charge threshold of \SI{120}{e} are shown in Figure~\ref{fig:cluster_charge_dis}.
The distributions are fitted with the convolution of a Gaussian and Landau function.
The MPV is \SI{1.42}{\kilo e} for both data and simulation, and the width of the Gaussian is \SI{0.21}{\kilo e}/\SI{0.22}{\kilo e} for data/simulation, respectively.
A good agreement between data and simulation is observed, as also indicated by the ratio of the two distributions displayed in the lower part of the figure.
While the MPV has been tuned to match data using the integration time of the simulation as discussed in Section~\ref{sec:simulation}, the agreement of the shapes indicates that the energy deposition in the relevant parts of the sensor as well as the collection of charge carriers is well-modeled by the simulation.
The data distribution exhibits some fluctuations owing to the low statistics of the sample.

\subsection{Cluster Size}
\begin{figure}[tbp]
	\centering
	\includegraphics[width=\columnwidth]{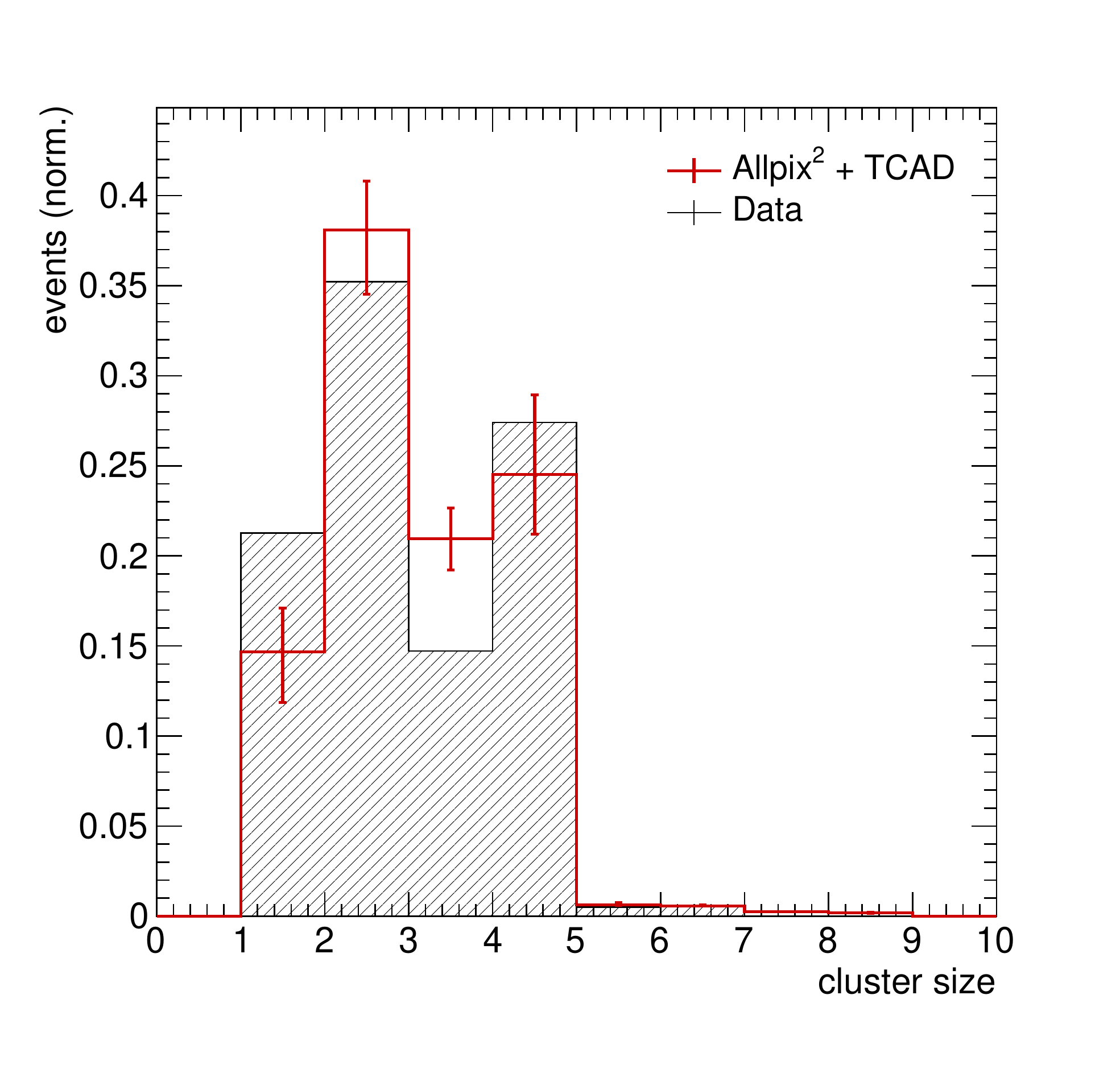}
	\caption{Cluster size distributions for experiment and simulation at a threshold of \SI{120}{e}.}
	\label{fig:cluster_size_dis}
\end{figure}

The distribution of the total cluster size at a threshold of \SI{120}{e} for simulation and experiment is presented in  Figure~\ref{fig:cluster_size_dis}.
Qualitatively, the distributions are in good agreement.
A possible source of the observed deviations for individual cluster sizes are uncertainties in the modeled electric field of the sensor as discussed in Section~\ref{sec:error}.

\begin{figure*}[tbp]
	\centering
	\includegraphics[width=\columnwidth]{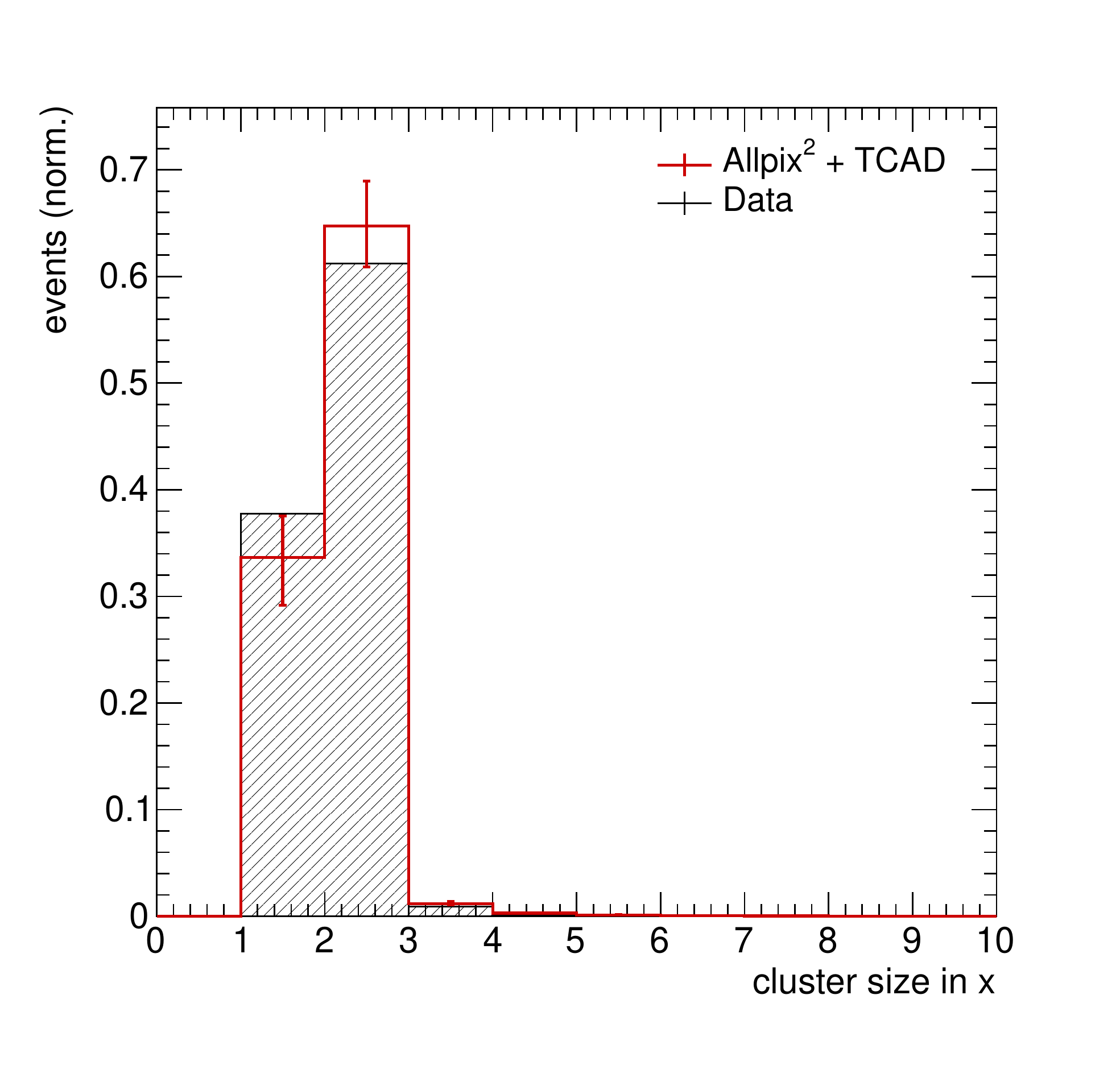}%
	\includegraphics[width=\columnwidth]{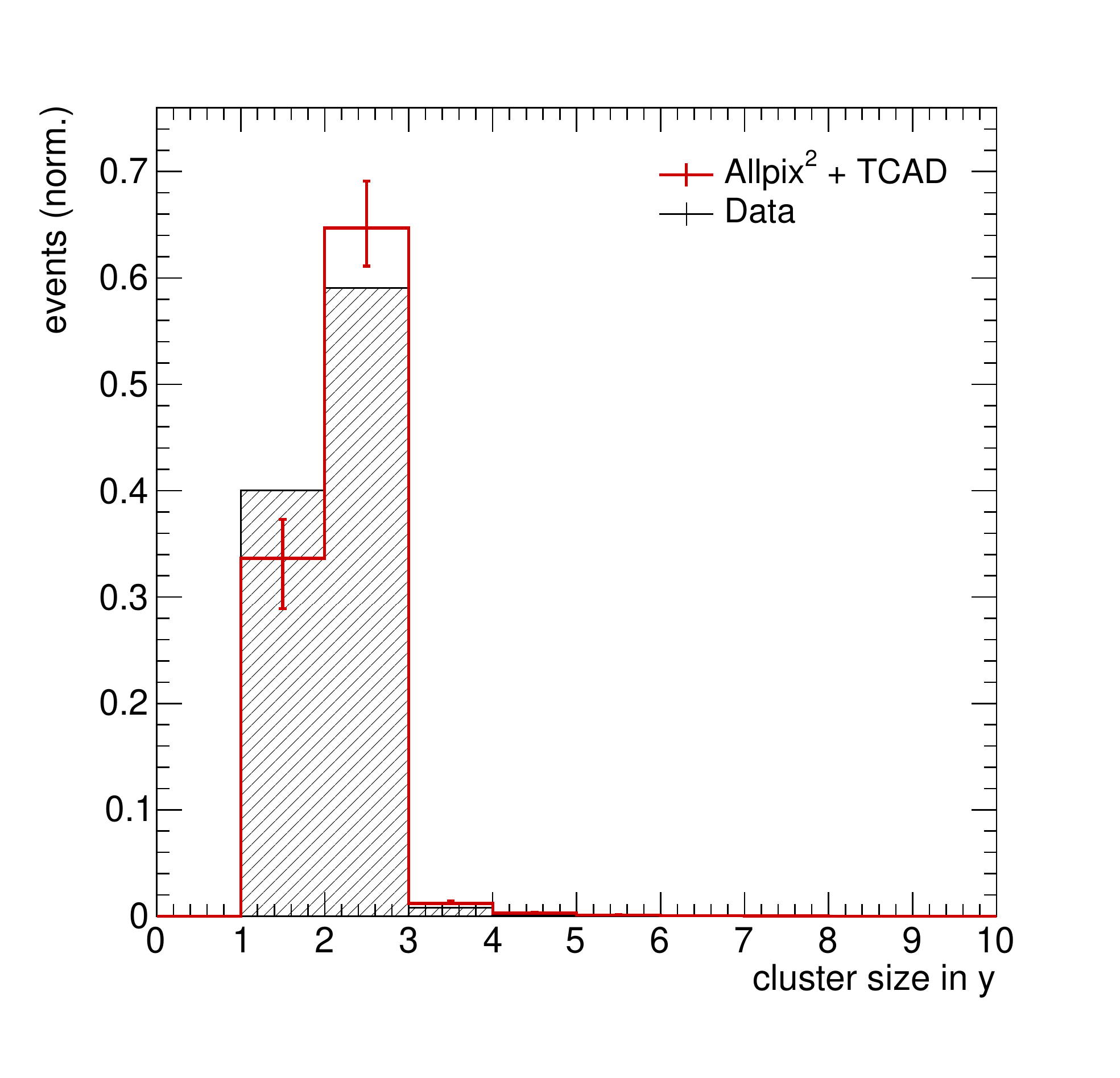}
	\caption{Cluster size projected in $x$ (left) and $y$ (right) at a threshold of \SI{120}{e} for data and simulation.}
	\label{fig:cluster_size_xy_dis}
\end{figure*}

The projection of the cluster size in $x$ and $y$, depicted in Figure~\ref{fig:cluster_size_xy_dis}, provides additional details about the charge sharing process.
Data and simulation agree well, but a small difference between the distributions in $x$ and $y$ can be observed in data despite the symmetry of the pixel cell layout.
It has been verified that this does not stem from a remaining misalignment in data by repeating the simulation with a sensor rotated around the $x$ axis by up to \SI{+-15}{\degree} in an attempt to reproduce the difference.
The deviation might be a result of the non-symmetric layout of the circuitry in the \emph{Investigator} pixel.
While the p-well structure has been designed to be fully symmetric in x and y, the layout of the circuitry placed in the p-wells is not symmetric, which is a possible source of the asymmetry.

\begin{figure}[tbp]
	\centering
	\includegraphics[width=\columnwidth]{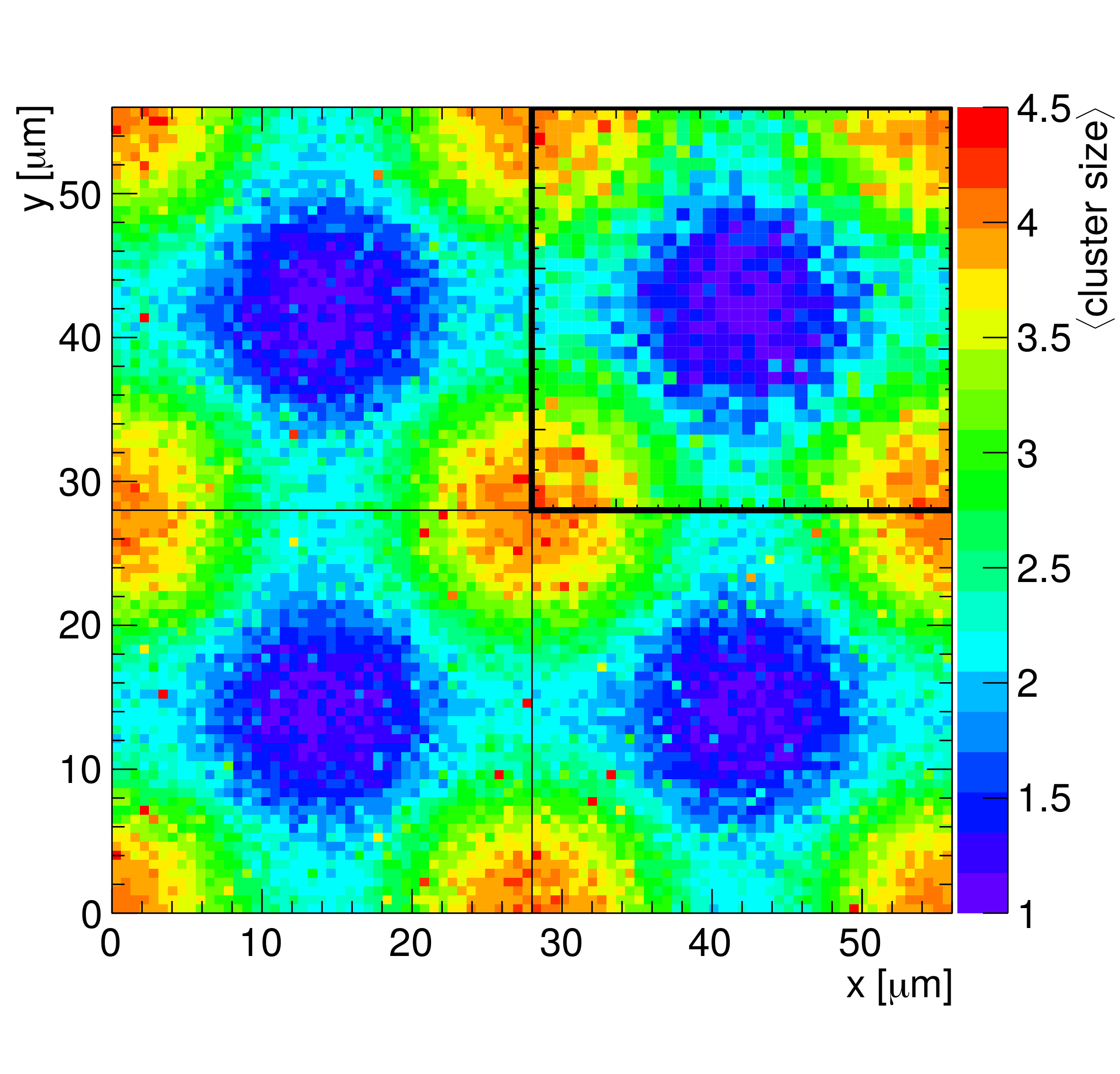}
	\caption{Intra-pixel representation of the cluster size for data and simulation at a threshold of \SI{120}{e}. Shown is an array of $2\times2$ pixel cells, with the top-right pixel displaying data, taken from~\cite{thesis-magdalena}, and the other pixels showing results from the simulation (color online).}
	\label{fig:cluster_size_map}
\end{figure}

The cluster size distribution is a precise measure for charge sharing as confirmed by the intra-pixel representation of the total cluster size presented in Figure~\ref{fig:cluster_size_map}.
For the simulation, the Monte Carlo truth information is exploited to produce a multi-pixel map indicating the mean cluster size as a function of the particle incidence position within the pixel cells.
Likewise, the reference track supplied by the beam telescope is used to obtain the particle incidence position for data. To increase statistics, data events from the full active matrix are folded into a single pixel cell, which is displayed in the upper-right quarter of Figure~\ref{fig:cluster_size_map}.

The largest clusters originate from the pixel corners since the low electric field between pixel implants results in a strong contribution from diffusion of charge carriers.
Single-pixel clusters, on the other hand, are almost exclusively produced if the incident particle traverses the sensor very close to the center of the pixel cell.

While the overall mean size distribution is faithfully reproduced in the simulation, minor discrepancies in the pixel corners are visible.
The transition from four to three-pixel clusters represented by the yellow regions is more apparent in simulation than in data.
The same holds true for the transition between two to three pixel clusters corresponding to the turquoise regions in Figure~\ref{fig:cluster_size_map}.
Particles penetrating the sensor at the corners of a pixel cell, for example, are more likely to give rise to clusters with size four in data compared to simulation.
This observation is in line with the higher number of clusters with size four in the cluster size distribution displayed in Figure~\ref{fig:cluster_size_dis}.
Moreover, the cluster size is particularly sensitive to a mis-modeling in the pixel corners as the diffusion of charge carriers to neighboring pixel cells is most likely if the incident particle enters the sensor at the corners between four cells.
Most notably, small modifications in the electric field in the pixel corners are capable of inhibiting or enhancing the motion of charge carriers to neighboring cells causing deviations in cluster size by up to two units as there are two cells directly adjacent to one corner.
As discussed in the previous section, the low field regions in the pixel corners are strongly influenced by the exact doping profile of the sensor.

\begin{figure}[tbp]
	\centering
	\includegraphics[width=\columnwidth]{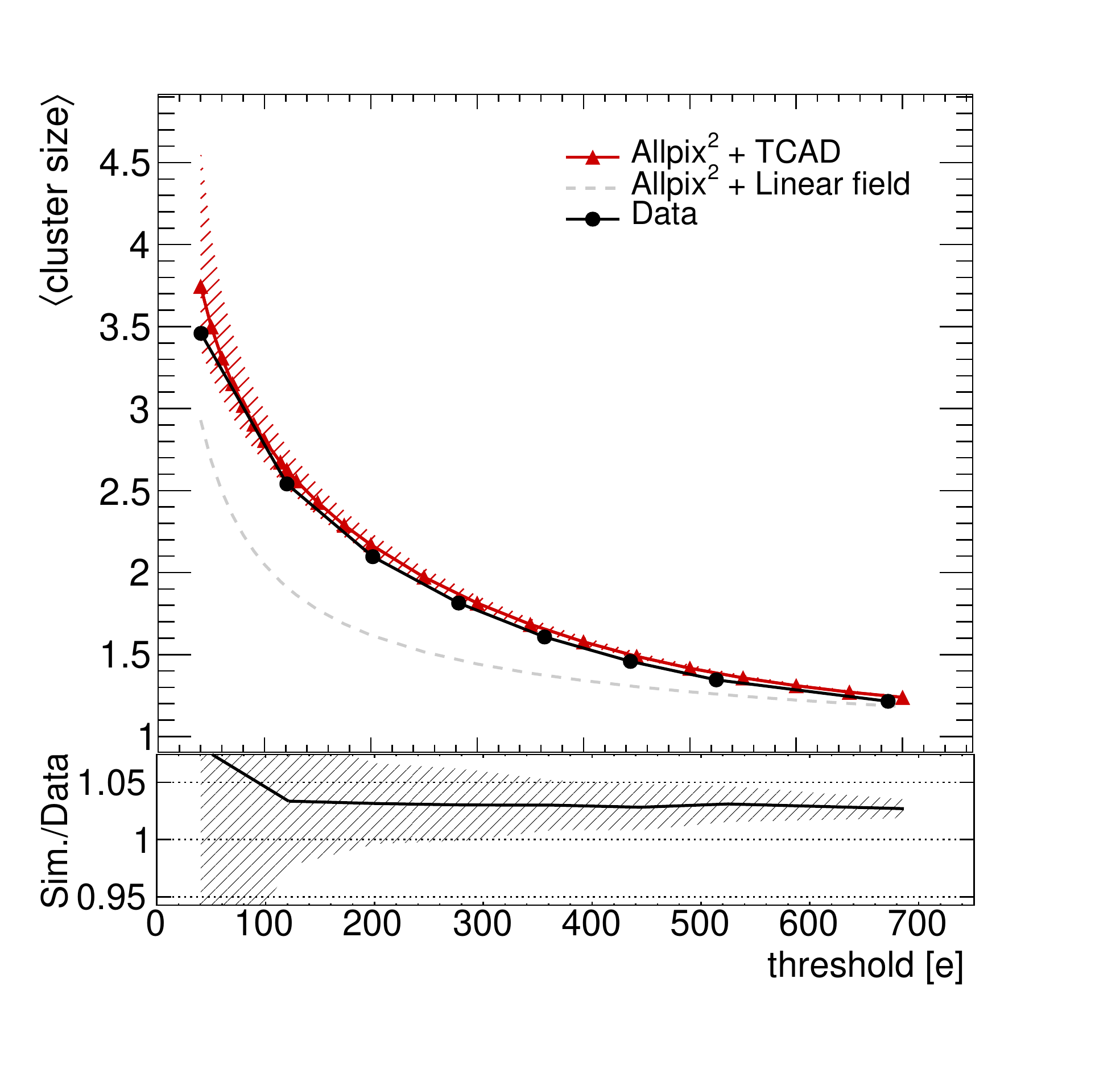}
	\caption{Mean cluster size as a function of the threshold, shown for experimental data as well as simulations with TCAD-modeled and linear electric fields. The hatched band represents the total uncertainty.}
	\label{fig:mean_cls_size}
\end{figure}

The mean cluster size has been studied as a function of the applied charge threshold.
Figure~\ref{fig:mean_cls_size} shows the curves for data and simulation.
In addition, a simulation with a linear electric field replacing the TCAD model in the epitaxial layer is plotted as a dashed line for comparison.
By increasing the threshold, the mean cluster size shifts to smaller values as individual pixels fall below the charge threshold.
Data and simulation match well down to very low thresholds, with a maximum deviation of about \SI{6}{\percent} at very low thresholds, while the simulation with a linear electric field produces incompatible results.
This deviation from the experimental results demonstrates the significance of a precise modeling of the electric field for this type of detector.
Similar results have been obtained for the mean projected cluster sizes along the $x$ and $y$ coordinates.

\begin{figure*}[tbp]
	\centering
	\includegraphics[width=\columnwidth]{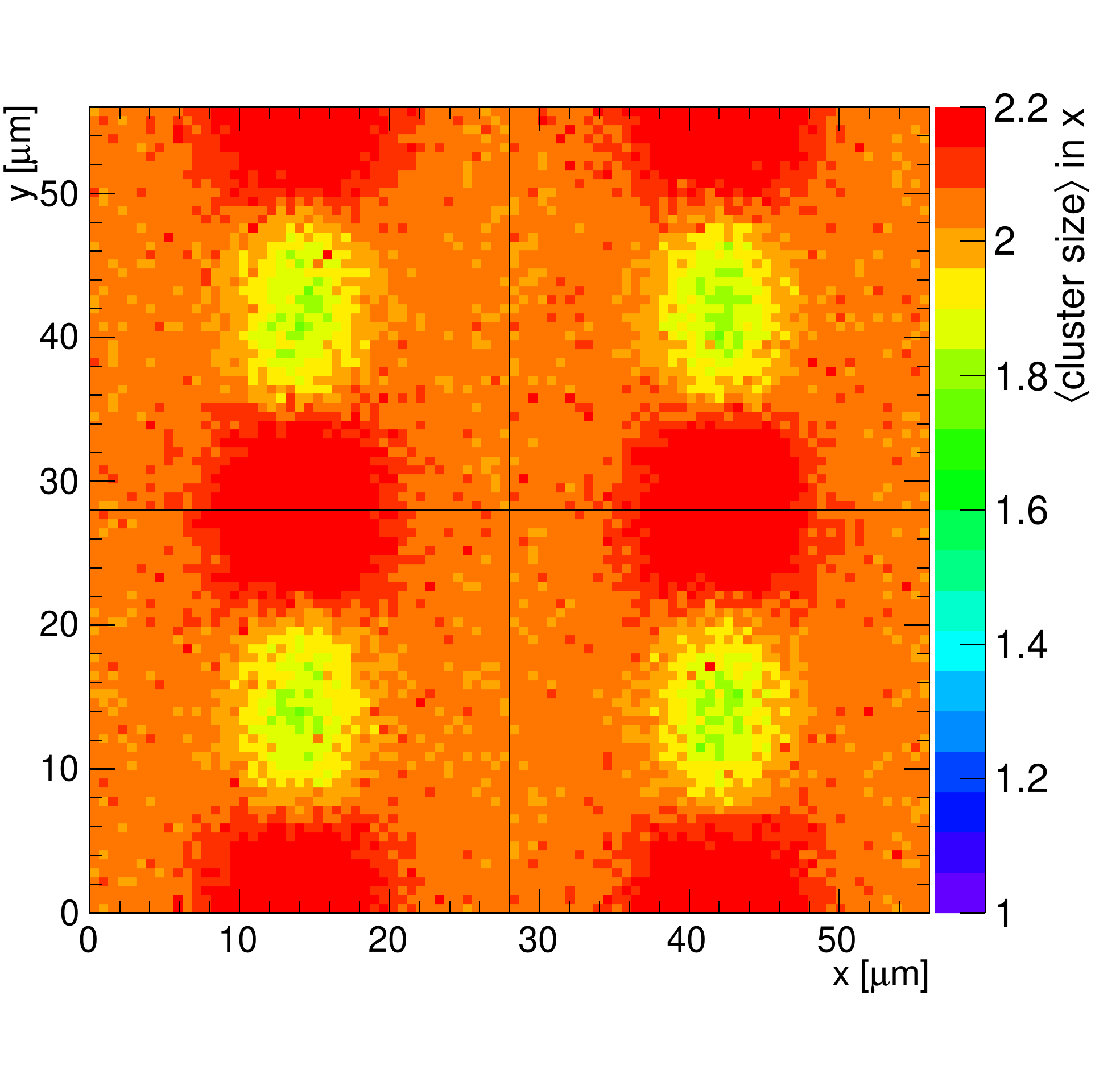}%
	\includegraphics[width=\columnwidth]{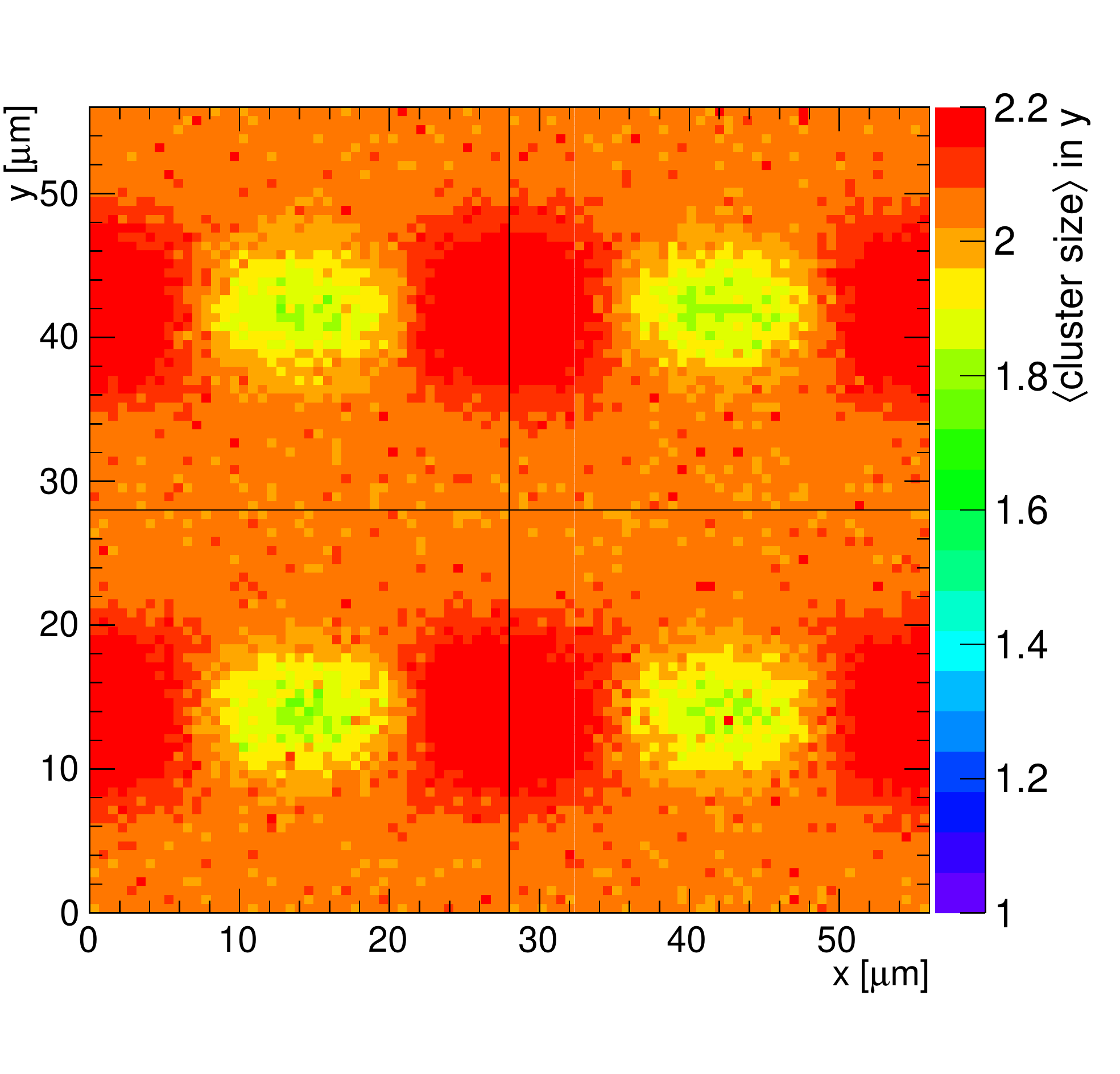}
	\caption{Intra-pixel representation of the mean projected cluster size as obtained from simulation. Shown are arrays of $2\times2$ pixel cells indicating the mean projected cluster size in $x$~(left) and $y$~(right) direction as a function of their relative position within the four pixel cells (color online).}
	\label{fig:cluster_size_xy_map}
\end{figure*}

Figure~\ref{fig:cluster_size_xy_map} displays $2\times2$ pixel maps of the mean projected cluster size in $x$ and $y$ as a function of the particle incidence position at a threshold of \SI{40}{e}.
Instead of the uniform bands along the respective coordinate expected for uncorrelated observables, eye-shaped structures reveal a correlation between charge sharing along the two dimensions caused by the inhomogeneous electric field and the bubble-shaped depletion region described in Section~\ref{sec:technology}.
The same effect is observed in data as demonstrated in~\cite{thesis-magdalena}.
With increasing threshold, charge sharing effects are suppressed and the correlation between the mean cluster size in $x$ and $y$ vanishes.


\section{Detector Performance}
\label{sec:performance}

Using the reconstructed cluster position and the Monte Carlo truth information from the primary particle, the performance of the CMOS detector is assessed in terms of spatial resolution and hit detection efficiency.
The results obtained from simulation are compared to data.

\subsection{Intrinsic Resolution}

\begin{figure}[tbp]
	\centering
	\includegraphics[width=\columnwidth]{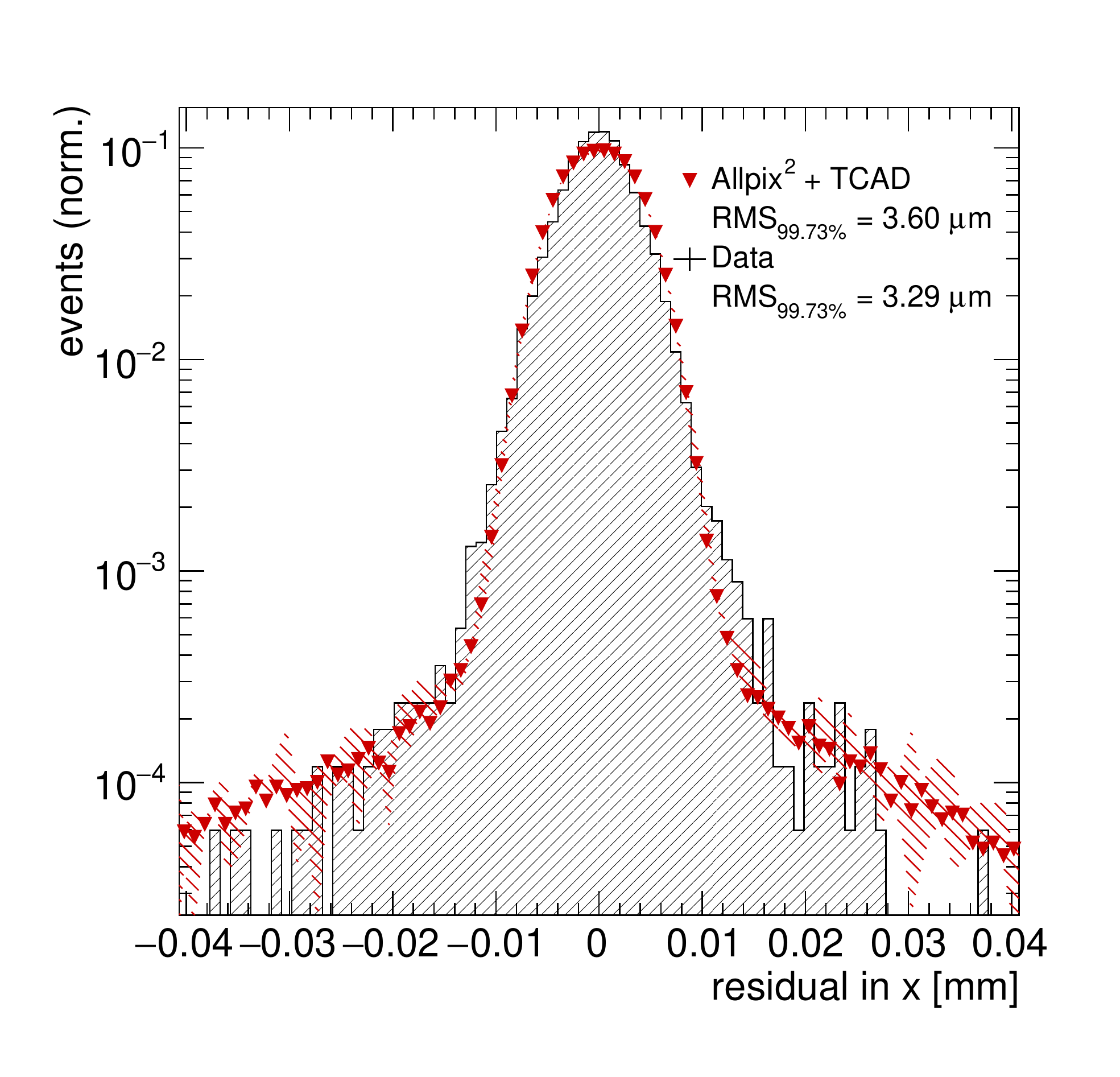}%
	\caption{Residuals in $x$ direction for data and simulation at a threshold of \SI{120}{e}. The hatched band represents the uncertainty on simulation.}
	\label{fig:residuals}
\end{figure}

Figure~\ref{fig:residuals} shows the residual in $x$, defined as the difference between the impact point of the incident particle obtained from the Monte Carlo truth and the reconstructed cluster position.
The width of the residual is obtained as the root mean square (RMS) of the distribution, evaluated for the central \SI{99.73}{\percent} of the histogram, equivalent to $\pm3\sigma$ of a Gaussian distribution, to match the definition used in the data analysis.
This allows the width of the distribution to be quantified independently from its shape while providing a statistically robust metric.

The spatial resolution is then calculated by quadratically subtracting the track resolution from the residual width, i.e.
\begin{align*}
\sigma = \sqrt{\textrm{RMS}_{\SI{99.73}{\percent}}^2 - \sigma_{\textrm{track}}^2} .
\end{align*}
The statistical uncertainty on the resolution is calculated using pseudo-experiments.
The number of entries in each bin of the residual distribution under consideration is smeared with a Poisson distribution with a mean equivalent to the original bin content.
The width obtained from the smeared histogram is stored, and the pseudo-experiment repeated \num{10000} times.
The statistical uncertainty on the residual width is then taken as the width of the resulting distribution and is propagated to the intrinsic resolution.

Using these definitions, resolutions in $x$ and $y$ of

$$\sigma_x = \SIERRA{3.60}{0.01}{+0.24}{-0.13}{\micro \meter}$$
$$\sigma_y = \SIERRA{3.57}{0.01}{+0.13}{-0.11}{\micro \meter}$$

have been achieved in simulation which is well below the value of pitch/$\sqrt{12} \approx \SI{8}{\micro m}$ expected without charge sharing.
It compares very well with the resolutions of \SI{3.29 \pm 0.02}{\micro \meter} and \SI{3.42 \pm 0.02}{\micro \meter} measured in data for $x$ and $y$ respectively.

\begin{figure*}[tbp]
	\centering
	\includegraphics[width=\columnwidth]{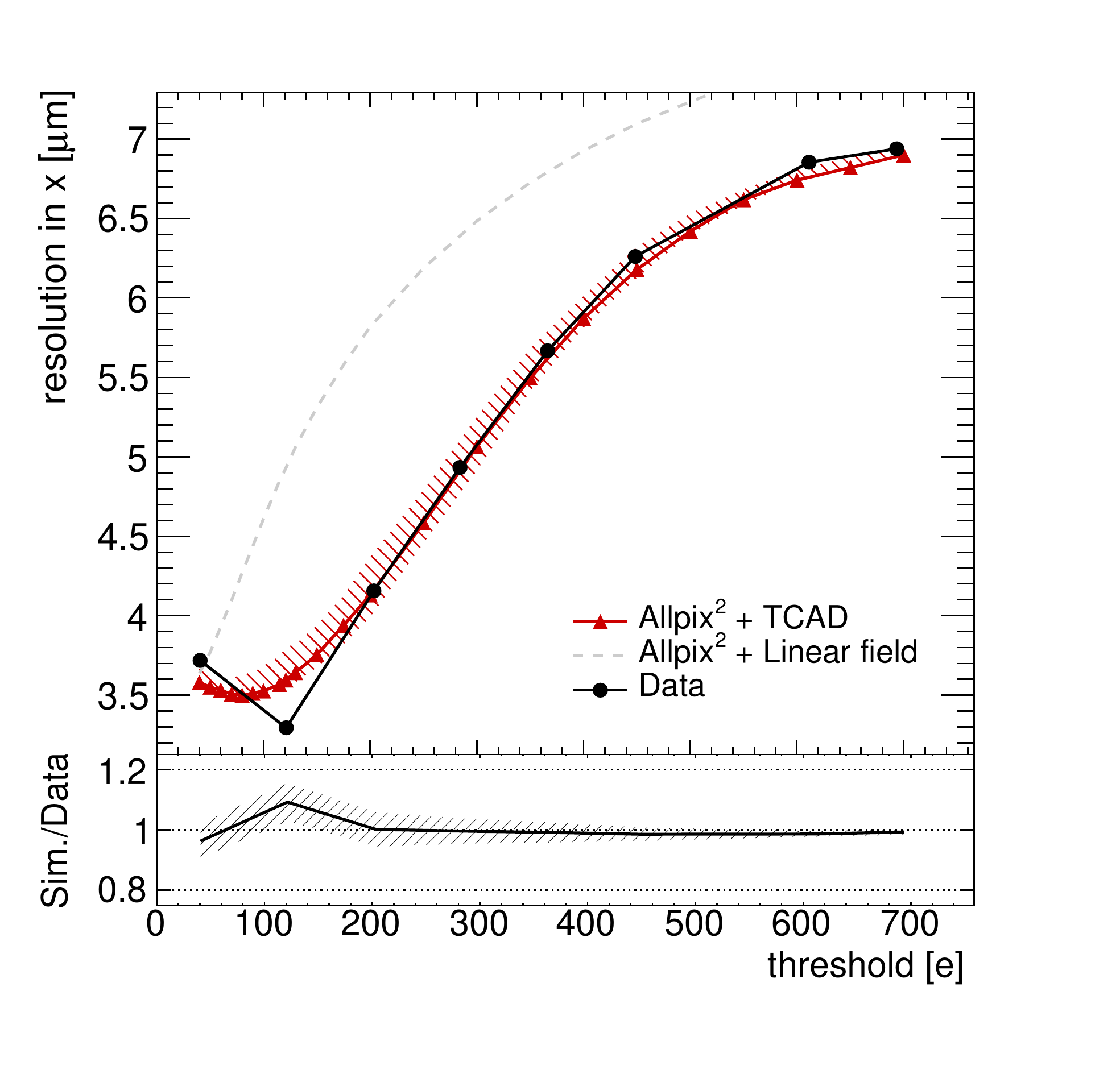}%
	\includegraphics[width=\columnwidth]{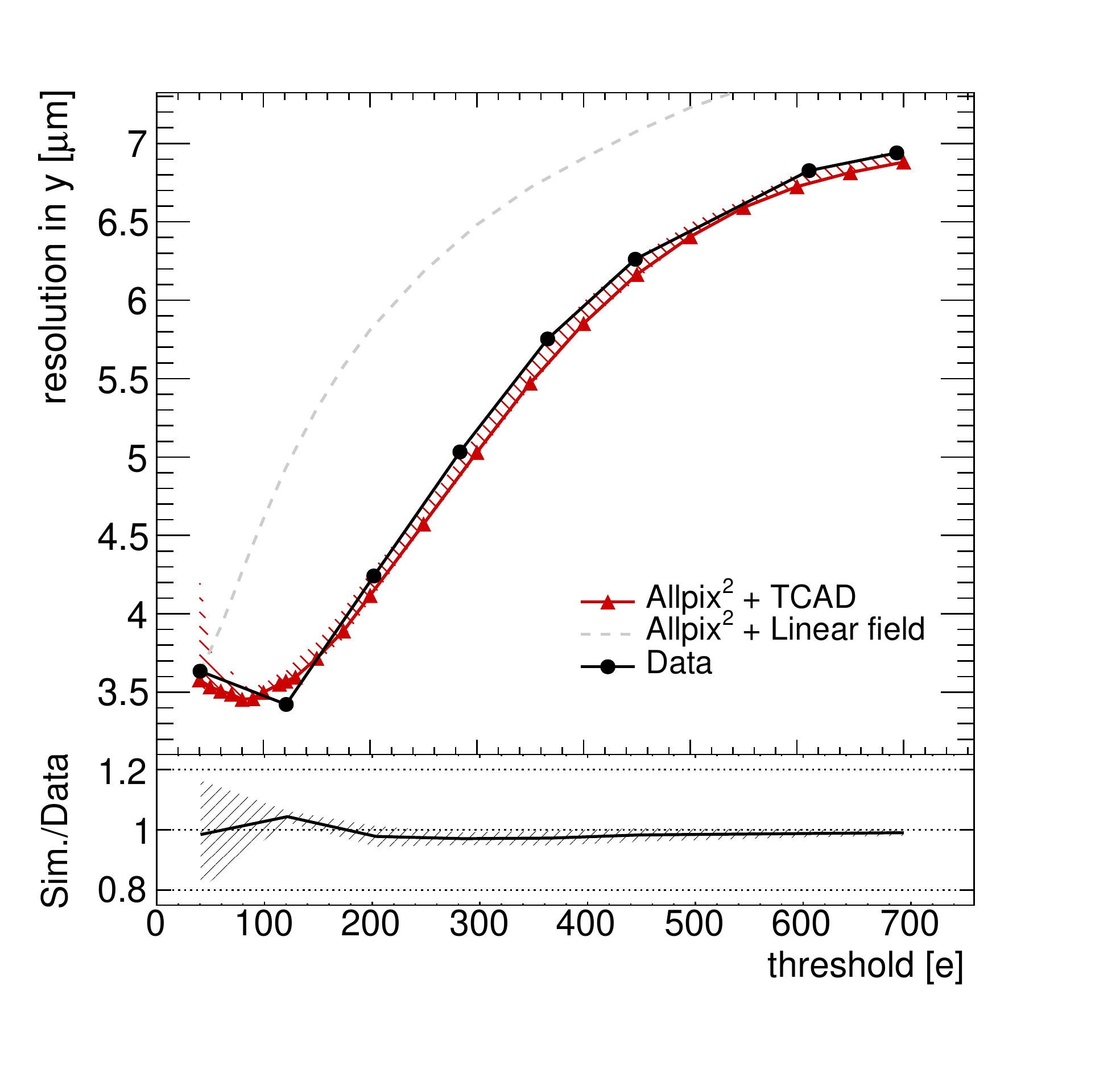}
	\caption{Spatial resolution in $x$~(left) and $y$~(right) direction as a function of the applied charge threshold, shown
for experimental data as well as simulations with TCAD-modeled and linear electric fields. The hatched band represents the total uncertainty.}
	\label{fig:mean_res_xy}
\end{figure*}

The resolution has been studied as a function of the charge threshold applied, shown in Figure~\ref{fig:mean_res_xy} for the $x$ and $y$ coordinates separately.
With increasing threshold, the information from pixels not passing the threshold is lost, leading to a deterioration of the position resolution.
The comparison of data with simulation shows a very good agreement down to a threshold of about \SI{150}{e}.
The discrepancy at lower thresholds is most likely to be a consequence of non-Gaussian noise in the data recorded with the analog prototype chip as well as a result of the simplification of charge carrier lifetimes described in Section~\ref{sec:simulation}.
The disagreement is of limited importance for practical purposes since a fully integrated sensor is likely to be operated at thresholds above \SI{150}{e}.

The dashed gray line in Figure~\ref{fig:mean_res_xy} again represents a simulation using a linear electric field as approximation, and the deviation from data suggests that this simplification leads to an inadequate description of the CMOS sensor response.

\subsection{Efficiency}

The efficiency of the detector is defined as the number of incident primary particles that can be matched to a reconstructed cluster divided by the total number of primary particles penetrating the detector.
A match between an incident particle and a reconstructed cluster is made, if the cluster is located within a radius of \SI{100}{\um} around the impact point of the incident particle, using the same matching criterion as applied to data.

The statistical uncertainty of the efficiency has been calculated by considering a counting experiment with two possible outcomes: either a matched or an unmatched primary particle track.
This results in an uncertainty of
\begin{align*}
\sigma_{\textrm{eff}} = \sqrt{\frac{p \cdot \left(1 - p\right)}{N}},
\end{align*}
where $p$ is the probability of a matched track while $N$ is the total number of experiments conducted.

\begin{figure*}[tbp]
	\centering
	\includegraphics[width=0.33\textwidth]{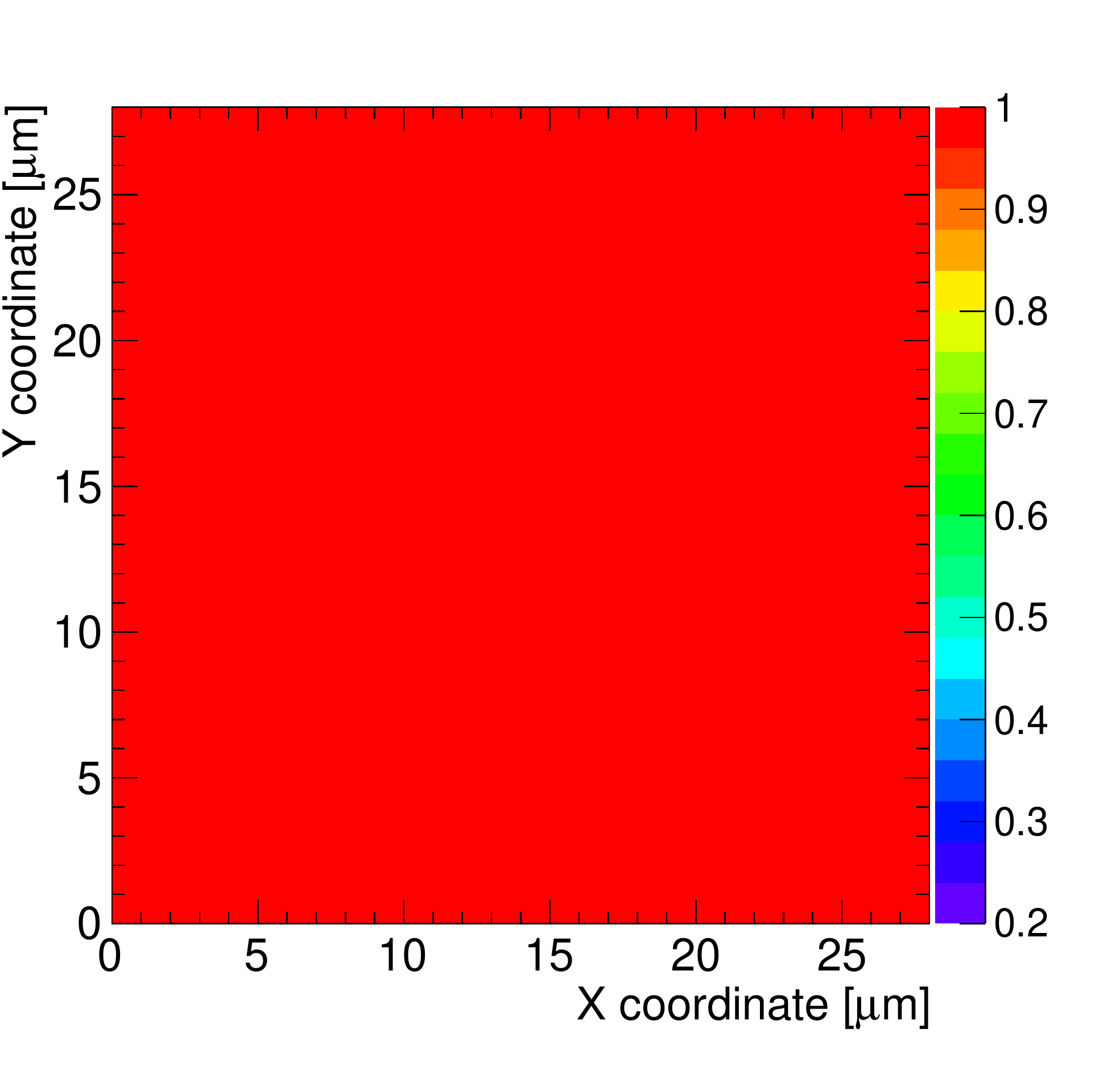}%
	\includegraphics[width=0.33\textwidth]{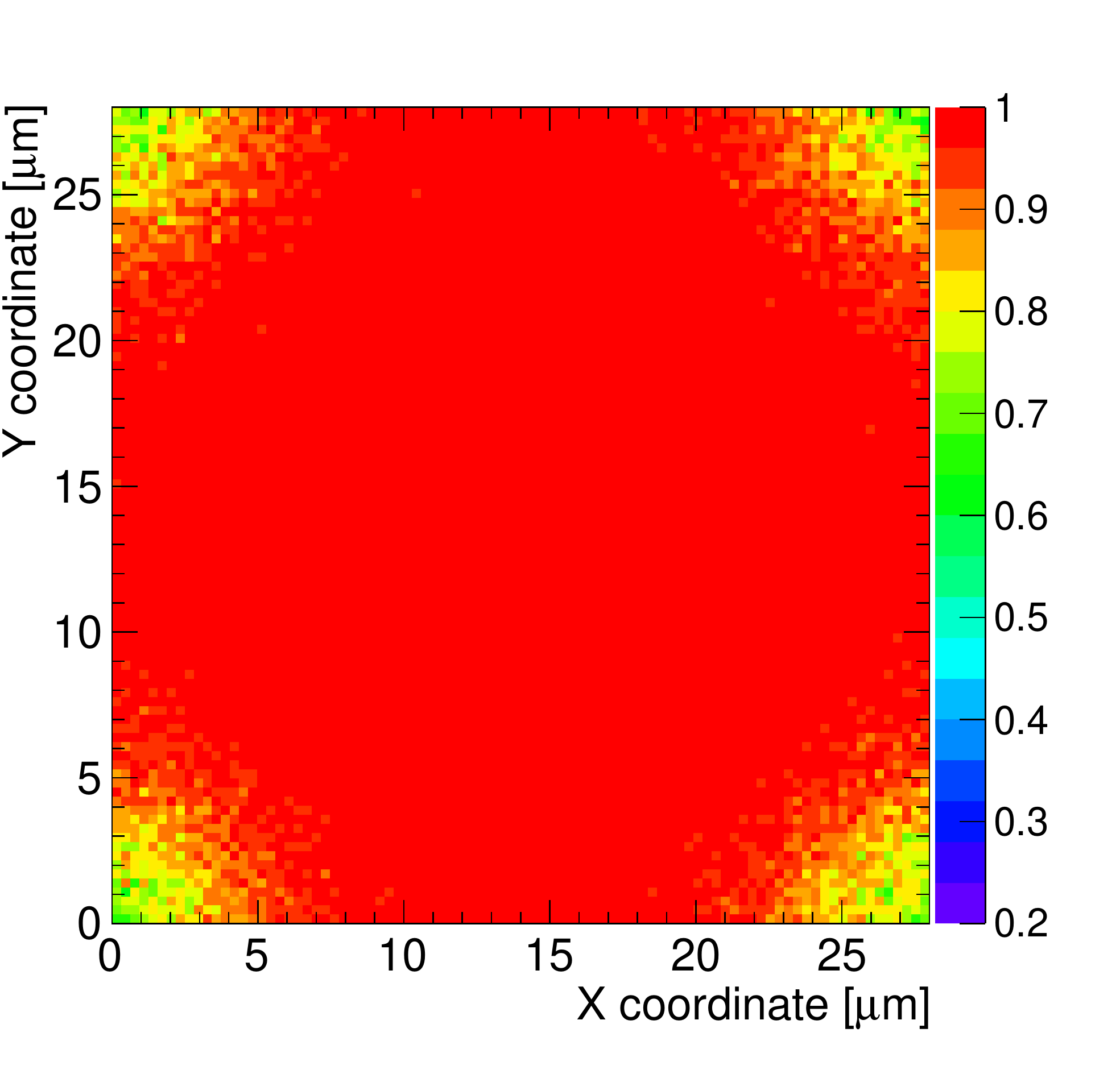}%
	\includegraphics[width=0.33\textwidth]{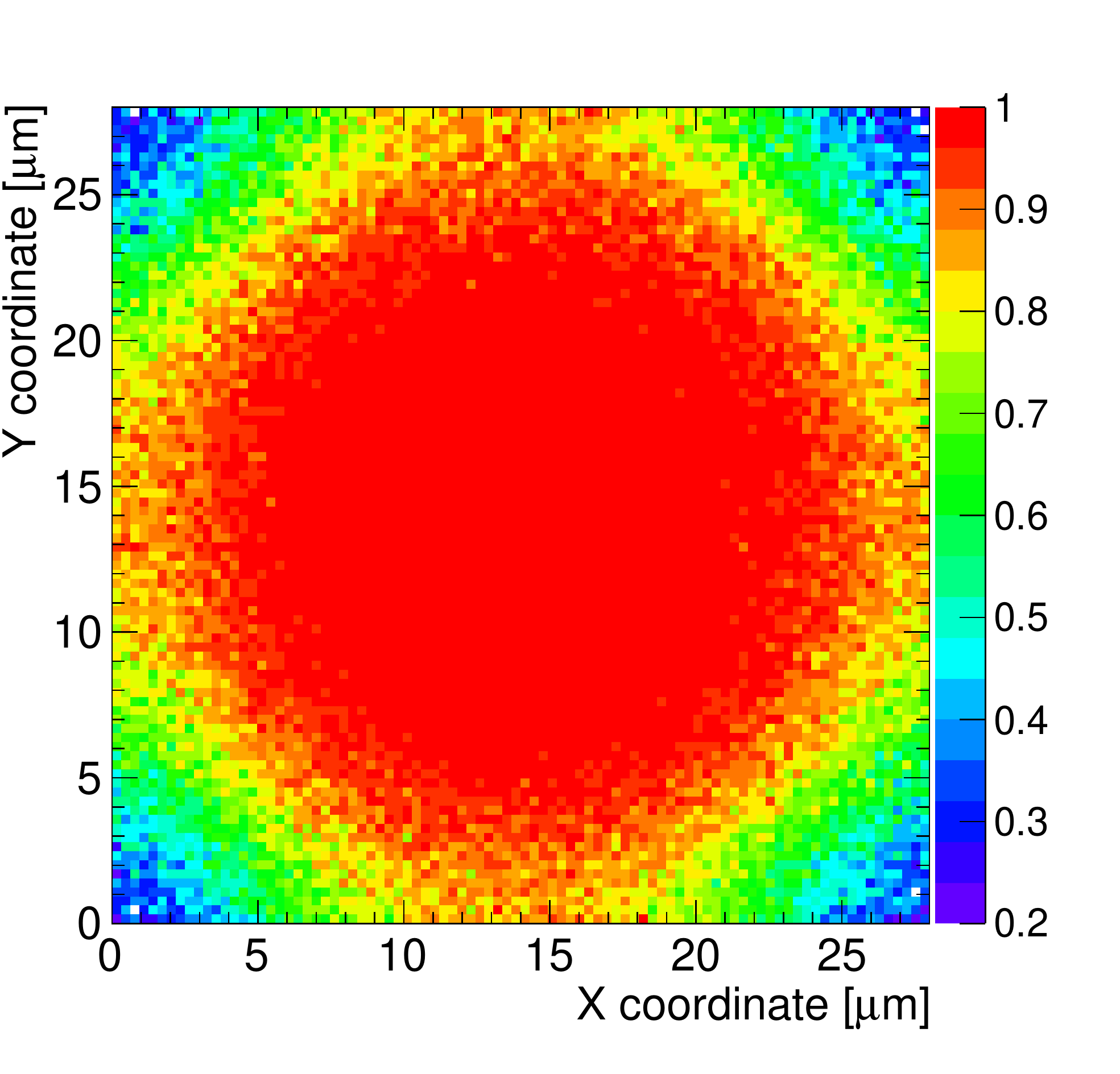}
	\caption{Efficiency obtained from simulations with TCAD-modeled electric field as a function of the impact position for charge thresholds of \SI{40}{e} (left), \SI{450}{e} (center) and \SI{700}{e} (right) for a single pixel cell (color online).}
	\label{fig:efficiency}
\end{figure*}

The efficiency obtained from simulation as a function of the particle impact position within a single pixel cell is displayed in Figure~\ref{fig:efficiency} for three different thresholds.

For the lower threshold of \SI{40}{e}, depicted in Figure~\ref{fig:efficiency}~(left), the simulation yields an overall efficiency of \SIERRS{99.95}{+0.05}{-0.23}{\percent}. The statistical uncertainty is of the order of \SI{1e-8}{}.
The remaining inefficiencies are evenly distributed throughout the pixel cell and arise from delta rays which pull the cluster center far away from the particle incidence point.
With increasing threshold, inefficiencies start to develop in the pixel corners, as these are the regions with the strongest charge sharing and the largest mean cluster size.
The overall hit detection efficiency at the threshold of \SI{450}{e} shown in Figure~\ref{fig:efficiency}~(center) decreases to about \SIERRS{97.62}{+0.13}{-0.58}{\percent}.
At the threshold of \SI{700}{e}, depicted in Figure~\ref{fig:efficiency}~(right), a pronounced inefficiency is observed, extending from the pixel corners into the pixel cell and leading to an overall efficiency of \SIERRS{85.96}{+0.53}{-1.02}{\percent}.

\begin{figure}[tbp]
	\centering
	\includegraphics[width=\columnwidth]{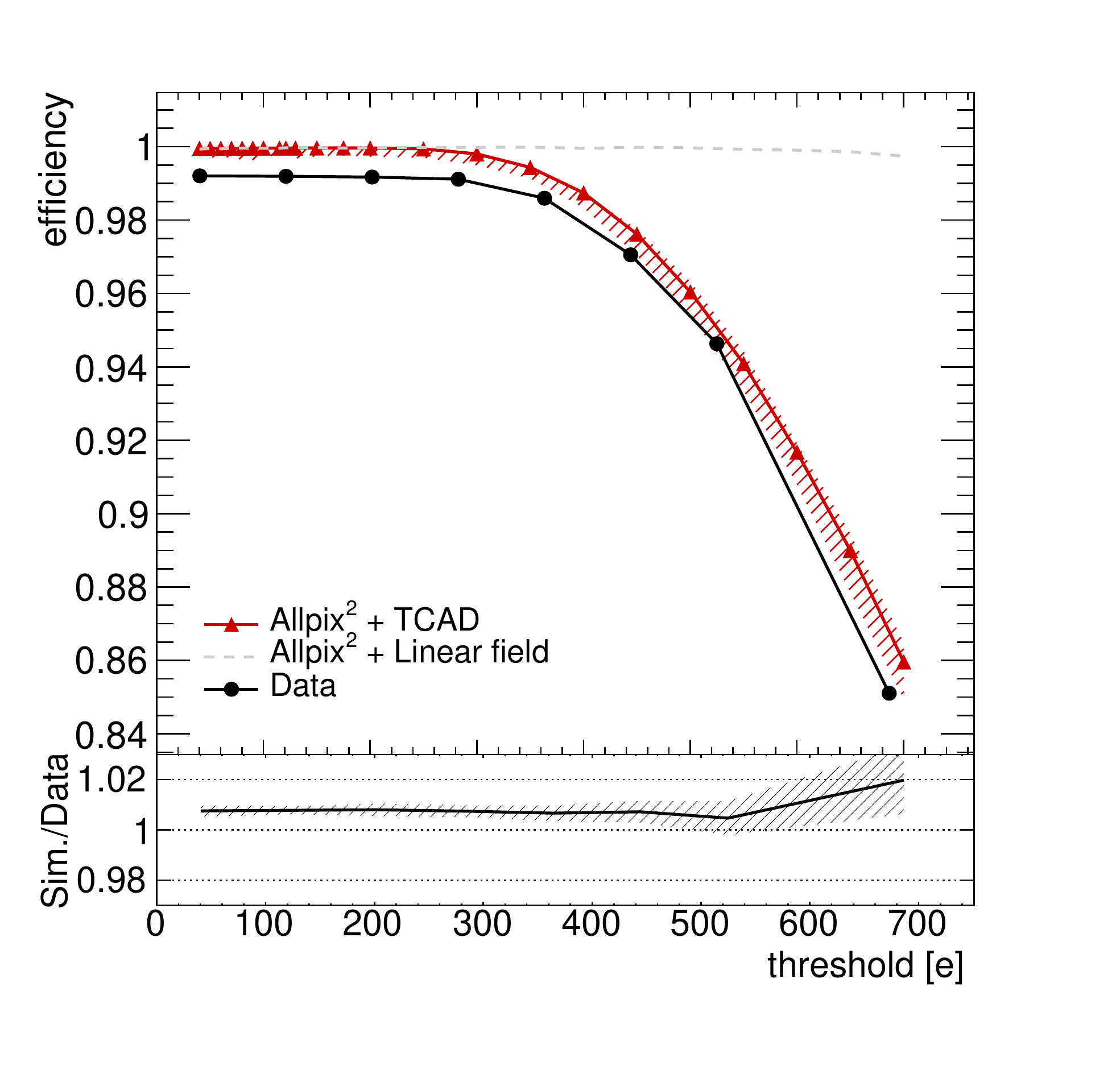}
	\caption{Efficiency as a function of the charge threshold, shown for experimental data as well as simulations with TCAD-modeled and linear electric fields. While the shape of the curve is well reproduced in simulation, a constant offset from data can be observed. The hatched band represents the total uncertainty.}
	\label{mean_eff}
\end{figure}

This decrease of efficiency can best be observed as a function of the charge threshold applied, as shown in Figure~\ref{mean_eff}.
While the shape of the curve observed in data is reproduced well, a constant offset to the measured values can be observed.
This difference can be attributed to fluctuations of the pedestal as well as inefficiencies in the data acquisition system which are not modeled in simulation.
The simulation using the linear electric field approximation is found to not correctly model the behavior observed at high threshold values.


\section{Summary \& Outlook}
\label{sec:summary}
In this paper, a combined 3D TCAD and Monte Carlo simulation of a CMOS pixel sensor with small collection electrode design, implemented in a high-resistivity epitaxial layer, has been presented.
The simulation combines the results of a three-dimensional electrostatic TCAD simulation with the stochastic description of energy deposition by Geant4 using the \apsq framework.
Visualizations of the charge carrier motion in the sensor produced by the simulation framework have been found to be helpful to qualitatively understand the sensor response.

The simulation results have been compared to measurements of a reference detector, recorded in a test-beam, and very good agreement has been observed after tuning the simulation to match the most probable value of the cluster charge measured in data.
The simplified charge transport model implemented in \apsq has been shown to be sufficiently precise to replicate the detector performance figures of merit such as efficiency and intrinsic resolution measured in data.

The implemented simulation setup for CMOS sensors will be used for further studies of similar detector prototypes and designs, including different sensor geometries and modified production processes aiming at a full lateral depletion of the epitaxial layer.

In future versions of the \apsq framework, a simulation of charge carrier recombination might be implemented, calculating the lifetime from the respective doping concentration as a function of their position within the sensor.
This would allow for an even more realistic description of the charge transport process and would remove the necessity of setting and tuning the integration time for underdepleted detectors.

Furthermore, the simulation could be extended to the detector performance in the timing domain by simulating the charge transport taking into account induced currents using the Shockley-Ramo theorem as possible with the latest version of the \apsq framework.

The presented combination of precise electric field modeling in TCAD and inclusion of statistical fluctuations is also interesting for the simulation of other silicon detector technologies with complex field configurations such as 3D sensors or LGADs.


\section*{Acknowledgements}
\label{sec:acknowledgements}
This work was carried out in the framework of the CLICdp Collaboration.
This project has received funding from the European Union's Horizon 2020 Research and Innovation programme under Grant Agreement no. 654168.


\section*{References}
\bibliography{bibliography}

\end{document}